%% file: arXiv-msc-uncertainty-vis.tex
\documentclass[10pt,journal,compsoc]{IEEEtran}

\ifCLASSOPTIONcompsoc
  \usepackage[nocompress]{cite}
\else	
  \usepackage{cite}
\fi

\ifCLASSINFOpdf
   \usepackage[pdftex]{graphicx}
  \graphicspath{{.}}
   \DeclareGraphicsExtensions{.pdf,.jpeg,.png,.jpg}
\else
\fi

\usepackage{amsmath}

\usepackage{algorithmic}

\usepackage{url}

\usepackage{mathtools}
\usepackage{amsfonts}
\usepackage{verbatim}
\usepackage{placeins}

\usepackage[dvipsnames]{xcolor}

\newcommand{\SM}{{Survival Map}} 
\newcommand{\PM}{{Probabilistic Map}}
\newcommand{\SSM}{{statistical summary maps}}

\newcommand{\para}[1]        {\vspace{2mm}\noindent{\textbf{#1}}}
\newcommand {\mm}[1] {\ifmmode{#1}\else{\mbox{\(#1\)}}\fi}

\newcommand{\Mspace}        {\mm{{\mathbb M}}}

\newcommand{\Rspace}        {\mm{{\mathbb R}}}

\newcommand{\grad}[1]     {{\nabla {#1}}}

\newcommand{\MS}        {\mm{\mathsf M}}

\newcommand{\PMap}        {\mm{\mathcal{P}}}
\newcommand{\SMap}        {\mm{\mathcal{S}}}
\newcommand{\CMap}        {\mm{\mathcal{C}}}

\begin{document}

\title{Uncertainty Visualization of 2D Morse Complex Ensembles Using Statistical Summary Maps}

\author{Tushar~Athawale,
        Dan~Maljovec,
        Chris~R.~Johnson,
        Valerio Pascucci,
        Bei Wang
\IEEEcompsocitemizethanks{\IEEEcompsocthanksitem T. Athawale, D. Maljovec, C. R. Johnson, V. Pascucci, and B. Wang are with Scientific Computing \& Imaging (SCI) Institute, University of Utah, Salt Lake City,
UT, 84112.
E-mails: \{tushar.athawale, maljovec, crj, pascucci, beiwang\}@sci.utah.edu}
}


\IEEEtitleabstractindextext{
\begin{abstract}
\input{sec-abstract}
\end{abstract}


}


\maketitle


\input{sec-introduction}

\input{sec-related-work}
\input{sec-background}

\input{sec-methods-pmap}

\input{sec-methods-smap}

\input{sec-results}
\input{sec-methods-noise}

\input{sec-conclusion}


\IEEEdisplaynontitleabstractindextext
\IEEEpeerreviewmaketitle

\section*{Acknowledgments}
This project is supported in part by NSF IIS-1513616, DBI-1661375,  and IIS-1910733; 
National Institute of General Medical Sciences of NIH  under grant P41 GM103545-18; and the Intel Parallel Computing Centers Program.



\input{arXiv-msc-uncertainty-vis.bbl}

\end{document}

%% file: sec-abstract.tex
Morse complexes are gradient-based topological descriptors with close connections to Morse theory. 
They are widely applicable in scientific visualization as they serve as important abstractions for gaining insights into the topology of scalar fields. 
Noise inherent to scalar field data due to acquisitions and processing, however, limits our understanding of the Morse complexes as structural abstractions.  
We, therefore, explore uncertainty visualization of an ensemble of 2D Morse complexes that arises from scalar fields coupled with data uncertainty. We propose \emph{statistical summary maps} as new entities for capturing structural variations and visualizing positional uncertainties of Morse complexes in ensembles. 
Specifically, we introduce two types of statistical summary maps -- the \emph{Probabilistic Map} and the \emph{Survival Map} -- to characterize the uncertain behaviors of local extrema and local gradient flows, respectively. We demonstrate the utility of our proposed approach using synthetic and real-world datasets.

%% file: sec-introduction.tex
\section{Introduction}
\label{sec:introduction}

Understanding the effects of data uncertainty on visualizations is one of the top research challenges~\cite{BonneauHegeJohnson2014, BrodlieAllendesLopes2012, JohnsonSanderson2003, PotterKnissRiesenfeld2010}. Uncertainty in visualizations cannot be averted due to noise inherent to data acquisitions, approximations during data processing, and the limitations of rendering devices~\cite{BrodlieAllendesLopes2012}.
The visualization of uncertainty can potentially improve our ability to reason about visualized data~\cite{HullmanQiaoCorrell2019}. 
A common practice to mitigate the effects of uncertainty is to combine  multiple simulations of a phenomenon (e.g.,~with varying parameters and different instruments) into an ensemble dataset; see~\cite{WangHazarikaLi2018} for a survey on ensemble visualization. 

\begin{figure}[!hb]
  \centering
    \includegraphics[width=0.98\columnwidth]{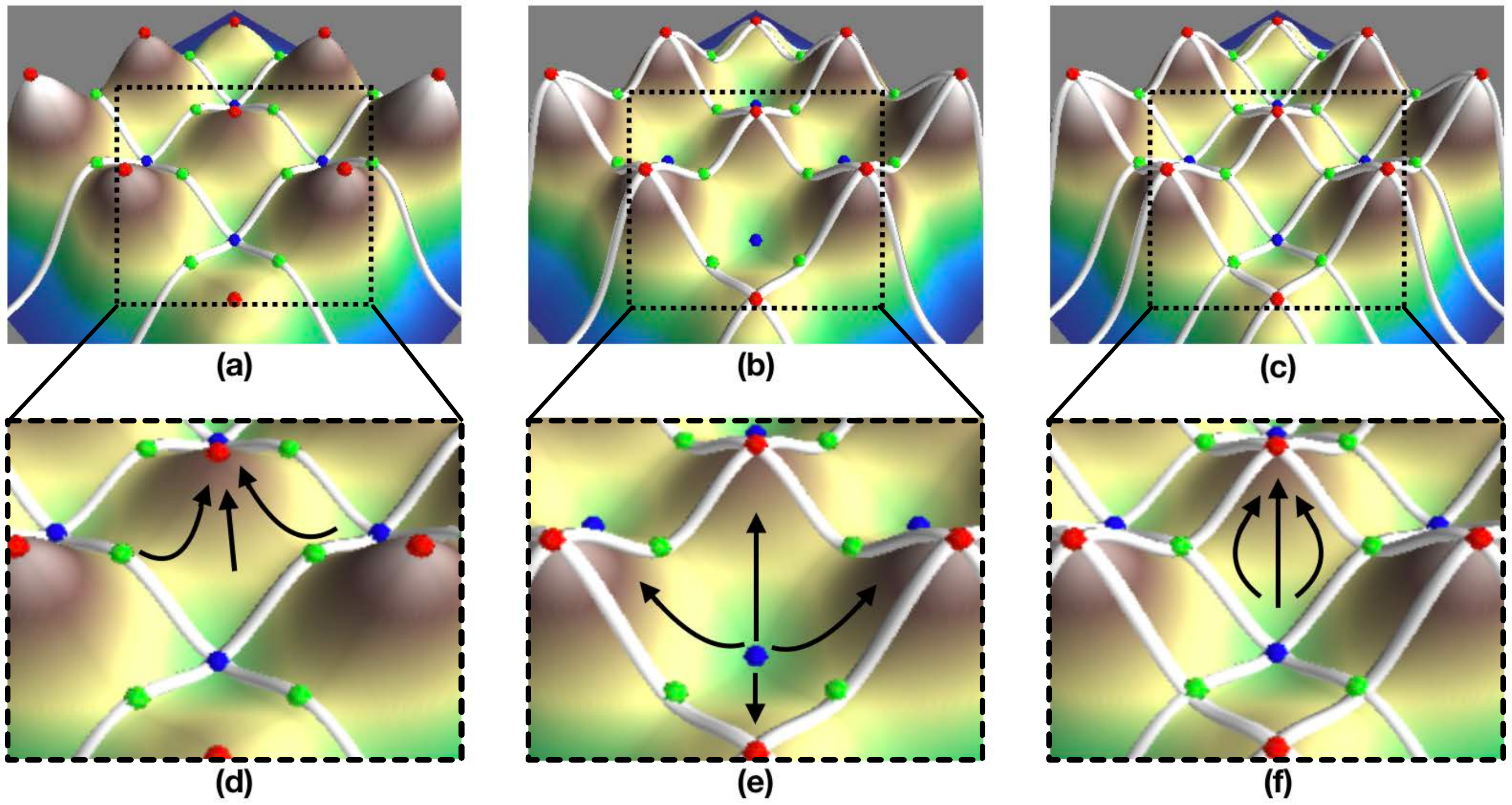}
    \vspace{-2mm}
    \caption{(a) Descending manifolds forming the Morse complex of $f$, (b) ascending manifolds forming the Morse complex of $-f$, (c) Morse-Smale complex of $f$. Local gradient flows in individual cells are depicted with the black arrows in the zoomed-in views (d)-(f).}
    \label{fig:msc}
\end{figure}

In this paper, we investigate the uncertainty in Morse complexes, an important type of topological descriptor, for an ensemble of 2D scalar fields. 
Morse complexes and Morse-Smale complexes are topological descriptors  based on Morse theory~\cite{Morse1925,Milnor1963} that provide abstract representations of the gradient flow behavior of scalar fields~\cite{Smale1961b,Smale1961}. 
Morse-Smale complexes have shown great utility in numerous scientific applications, from identifying burning regions in combustion experiments \cite{BremerWeberPascucci2010} to counting bubbles in mixing fluids \cite{LaneyBremerMascarenhas2006}. 
They also appear in partition-based regression~\cite{GerberRubelBremer2012,MaljovecWangRosen2016} and statistical inference~\cite{ChenGenoveseWasserman2017}.

Morse complexes~\cite{EdelsbrunnerHarerZomorodian2003} are the building blocks for Morse-Smale complexes. 
Given a Morse function $f$ defined on a manifold $\Mspace$, $f: \Mspace \to \Rspace$, 
the Morse complex of $f$ decomposes the domain into cells, referred to as \emph{descending manifolds}, where points in the same cell have their gradient flows terminate at the same local maximum. 
The Morse complex of $-f$ decomposes the domain into cells, referred to as \emph{ascending manifolds}, where points in the same cell have their gradient flows originate from the same local minimum. 
If the ascending and descending manifolds intersect transversally, the set of intersections creates the Morse-Smale complex of $f$, which partitions the domain into cells with uniform gradient behavior. 
Fig.~\ref{fig:msc} illustrates the Morse and Morse-Smale complexes of a 2D height function $f$. 
In particular, for the Morse complex in Fig.~\ref{fig:msc}a, $0$-cells are the critical points of $f$ (red for local maxima, blue for local minima, and green for saddles), $1$-cells are integral lines (in white) connecting the critical points, and $2$-cells are connected regions in the domain separated by $1$-cells (see Sec.~\ref{sec:background} for details).  

Morse and Morse-Smale complexes have been extensively studied under both piecewise-linear (PL) and combinatorial settings (see Sec.~\ref{sec:relatedwork}). However, visualization of Morse and Morse-Smale complexes in the face of uncertainty remains challenging. By uncertainty we mean information about their accuracy, confidence, and variability~\cite{BonneauHegeJohnson2014}. In terms of accuracy, Gyulassy et al.~\cite{GyulassyBremerPascucci2012} have introduced algorithms that improve upon the geometric quality of Morse-Smale complexes. 
Their algorithms are shown to produce the correct results on average, and the standard deviation approaches zero with increasing mesh resolution. 
In terms of variability, Thompson et al.~\cite{ThompsonLevineBennett2011} have briefly mentioned a Monte Carlo sampling method to quantify variations in the boundaries of Morse complexes. 

Motivated by limited prior work in encoding uncertainty of topological descriptors~\cite{HeineLeitteHlawitschka2016,YanWangMunch2019}, we study the uncertainty in Morse complexes for an ensemble of 2D scalar fields under a particular noise model. 
Suppose $n$ ensemble members are given as scalar functions defined on a shared 2D domain, $f_1, \cdots, f_n:  \Mspace \to \Rspace$, where $ \Mspace \subset \Rspace^2$. We study an ensemble of Morse complexes $\MS_1, \cdots, \MS_n$ computed from these functions.  
We assume that each ensemble member $f_i$ is drawn from a  \emph{distribution} that is concentrated around a ground truth function $f$, i.e., $f_i(x) \sim f(x) \pm  \epsilon_i(x)$ for any $x \in \Mspace$. 
The specific models of noise $\epsilon_i(x)$ can be quite general (parametric or nonparametric), but we assume $\epsilon_i(x)$ to be \emph{upper bounded} by  half of the \emph{persistence} of the smallest topological feature of the ground truth function $f$ (see Sec.~\ref{sec:background}, Sec.~\ref{sec:probabilisticMap}, and Sec.~\ref{sec:noiseModel} for details).

In this work, we propose {\SSM} as new entities for visualizing structural variations of Morse complexes. 
Such structural variations encompass positional uncertainties of $2$-cells of an ensemble of Morse complexes $\MS_1, \cdots, \MS_n$ that arises from a common domain $\Mspace$ and a ground truth function $f$. 
We introduce two types of {\SSM}, the {\PM}, $\PMap: \Mspace \to \Rspace^l$ (where $l$ is the number of local maxima of $f$), and the {\SM}, $\SMap: \Mspace \to \Rspace$, to be utilized in  uncertainty visualization.

\begin{figure}[!ht]
  \centering
    \includegraphics[width=0.48\textwidth]{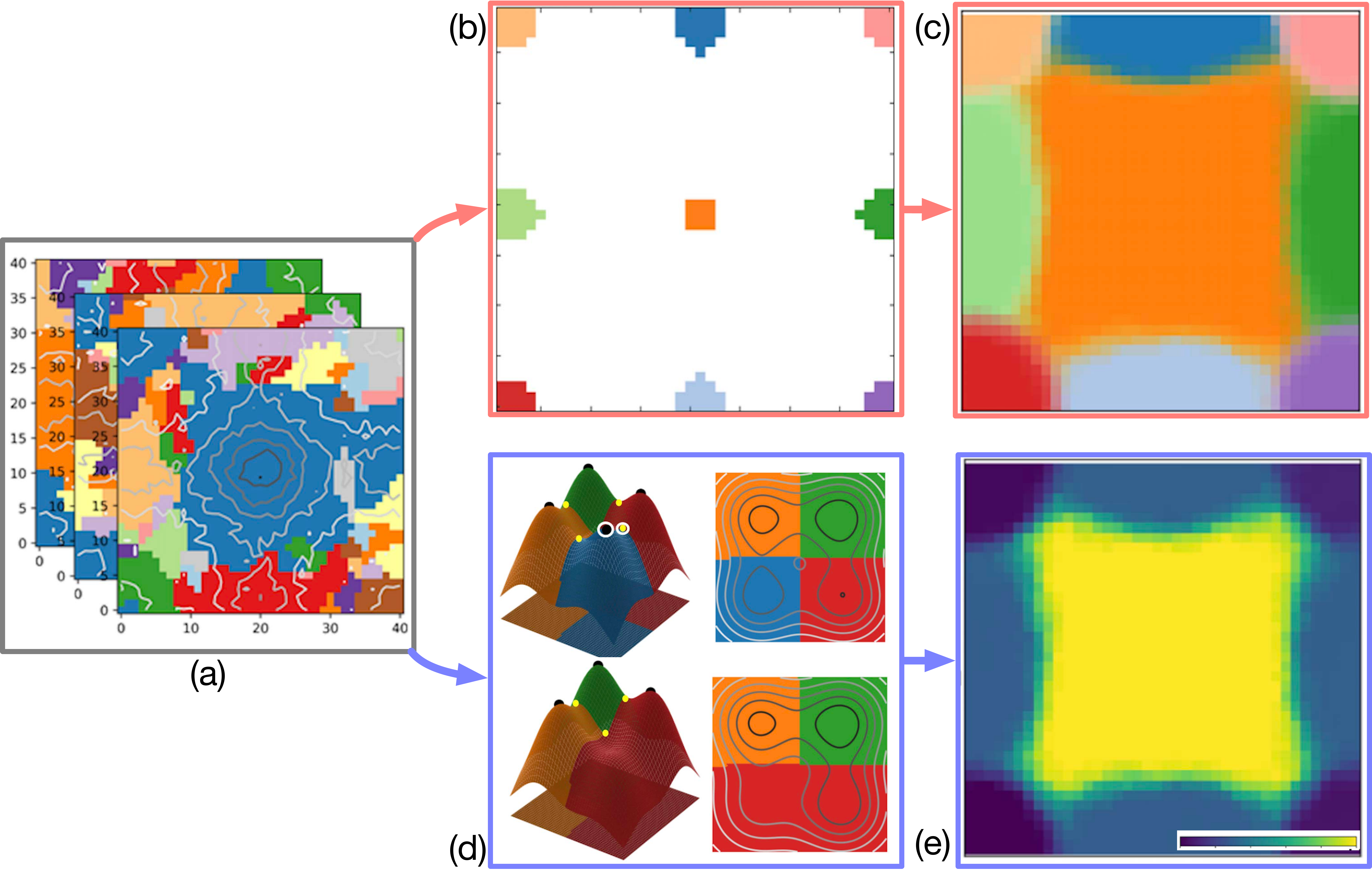}
    \vspace{-2mm}
    \caption{An overview of our computational pipeline. (a) An input  ensemble of 2D Morse complexes, (b) clusters of mandatory local maxima, (c) a {\PM} for the input ensemble,  (d) persistence simplification of each ensemble member, (e) a {\SM} for the input ensemble.}   
    \label{fig:pipeline}
\end{figure}

\para{Overview.} 
We begin with an overview of our computational pipeline for deriving {\SSM}, as illustrated in Fig.~\ref{fig:pipeline}. 

Given an ensemble of 2D scalar fields, we compute an ensemble of Morse complexes $\MS_1, \cdots, \MS_n$. 
For our first approach, we study positional uncertainties of local maxima across the ensemble to derive a {\PM}.
We first extract (clusters of) mandatory local maxima across the ensemble using techniques developed by David et al.~\cite{DavidJosephJulien2014}. 
Specifically, we demonstrate that mandatory local maxima have a one-to-one correspondence with the local maxima of the ground truth function $f$ under our noise model (Fig.~\ref{fig:pipeline}b and Sec.~\ref{sec:noiseModel}). 
We assign a label to each mandatory local maxima (and, equivalently, to each local maxima of $f$); let $[l] = \{1, 2, \cdots, l\}$ denote the set of labels.  
Second, for each point in the domain, we compute a probability distribution of its cluster membership across the ensemble. 
That is, fix an ensemble member $\MS_i$, we trace the ascending integral line of each point $x \in \Mspace$ toward its destination, a local maximum $y \in \Mspace$, and assign to $x$ the label of $y$ as its cluster membership; let $\alpha_i: \Mspace \to [l]$ denote such an assignment. 
The {\PM} $\PMap: \Mspace \to \Rspace^{l}$ is defined as a discrete probability distribution of $\{\alpha_1(x), \alpha_2(x), \cdots, \alpha_n(x))\}$ for each $x \in \Mspace$. 
We visualize the {\PM} using color blending in Fig.~\ref{fig:pipeline}c, where each color represents a distinct label of a mandatory local maximum; see Sec.~\ref{sec:probabilisticMap} for details. 

For our second approach, we quantify structural deviations in local gradient flows across the ensemble via a {\SM}. 
Specifically, we study directional changes of gradient flows as a result of persistence simplification~\cite{EdelsbrunnerLetscherZomorodian2002}.
For a fixed ensemble member $\MS_i$, we first apply a hierarchical  persistence simplification (Fig.~\ref{fig:pipeline}d) of $\MS_i$ using persistence as a scale parameter. 
We assign a \emph{survival measure} for each point $x \in \Mspace$ based on how frequently it changes its local gradient flows during the  simplification process. 
The less frequently $x$ changes its gradient directions, the greater is its survival measure, and vice versa. 
In other words, the survival measure quantifies the \emph{survivability} of consistent flow behaviors. 
Let $\beta_i: \Mspace \to \Rspace$ denote such an assignment of the survival measure. 
The {\SM} $\SMap: \Mspace \to \Rspace$ is defined to be the average value of survival measures across the ensemble $\{\beta_1(x), \beta_2(x), \cdots, \beta_n(x)\}$ for each $x \in \Mspace$; that is, $\SMap(x) = \frac{1}{n}\sum_{i=1}^{n} \beta_i(x)$.
We visualize the {\SM} using a heat color map (blue means low and yellow means high survivability value); see Sec.~\ref{sec:survivalMap} for details.
  
\para{Contribution.}
In summary, given a 2D Morse complex ensemble: 
\begin{itemize} 
\item We exploit mandatory local maxima~\cite{DavidJosephJulien2014} that capture positional uncertainties among local maxima across the ensemble to derive a \emph{\PM}. 
\item We employ information obtained during persistence simplification~\cite{EdelsbrunnerLetscherZomorodian2002} of each ensemble member that characterizes structural variations among local gradient flows to derive a \emph{\SM}. 
\item We apply various uncertainty visualization techniques, such as interactive probability queries~\cite{PotterKirbyXiu2012}, to our {\PM} and {\SM} for understanding the Morse-complex structural uncertainty in synthetic and real-world datasets. 
\end{itemize}

%% file: sec-related-work.tex
\section{Related Work}
\label{sec:relatedwork}

\para{Representations of Morse-Smale Complexes.}
Morse and Morse-Smale complexes are defined for functions on smooth $d$-manifolds. 
Moving from the smooth category to the discrete category requires considerable effort to ensure structural integrity and to simulate  differentiability~\cite{EdelsbrunnerHarerZomorodian2003}. 
In general, Morse-Smale complexes can be represented explicitly or implicitly~\cite{GuntherReininghausSeidel2014}. 
The first, an explicit representation, is computed in  2D~\cite{EdelsbrunnerHarerZomorodian2003} and 3D~\cite{EdelsbrunnerHarerNatarajan2003} for piecewise linear (PL) functions defined on triangulated domains. 
The $1$-skeleton ($0$-cells and $1$-cells) of a Morse-Smale complex is represented as a graph, referred to as the Morse-Smale graph, which connects critical points (as nodes) with separatrices (as edges). 
The second, an implicit representation, originates from Discrete Morse theory~\cite{Forman1998, Forman1998a, Forman1999, Forman2002} where a Morse-Smale complex is implicitly represented by a combinatorial gradient field~\cite{GuntherReininghausSeidel2014}.
The simplification of Morse-Smale complexes works differently depending on their representations; see~\cite{GuntherReininghausSeidel2014} for a detailed investigation.  
Many algorithmic efforts have focused on practical and efficient  computations using either of these representations (e.g.,~\cite{GyulassyNatarajanPascucci2007, GyulassyBremerPascucci2008, ReininghausLowenHotz2011}). 
We present our results for the explicit representations of Morse complexes, although our methods do not depend upon the choice of the  representation. 
Note that in image analysis, the watershed algorithm~\cite{BeucherMeyer1993} is analogous to the computation of Morse complexes in low dimensions.

\para{Uncertainty visualization of critical points and gradient fields.}
For a scalar function, its critical points and induced gradient field characterize the structure of its corresponding Morse complex. 
A few recent works have focused on data uncertainty and its effects on the critical points and gradient fields. 
Mihai and Westermann~\cite{MihaiWestermann2014} have proposed likelihood visualizations of the critical points for an uncertain scalar field. 
H\"{u}ttenberger et al.~\cite{HuettenbergerHeineCarr2013} have exploited the idea of Pareto optimality for predicting the positions of local extrema for multifield data. 
G{\"u}nther et al.~\cite{DavidJosephJulien2014} have devised mandatory critical regions as a way to segment the domain of uncertain data, where at least one critical point of an  unknown underlying function is guaranteed to exist within a mandatory critical region. 
Favelier et al.~\cite{FavelierFarajSumma2019} have developed  persistence-based clustering of ensemble members followed by mandatory critical regions for visualizing positional uncertainties of critical points. 
In this work, we leverage the idea of mandatory critical regions in our {\PM} (Sec.~\ref{sec:probabilisticMap}). 

Pfaffelmoser et al.~\cite{PfaffelmoserMihaiWestermann2013} have analyzed the variability in gradient fields induced by uncertain scalar fields; where gradients are computed using the notion of \emph{central differences}. 
 Otto et al.~\cite{OttoGermerHege2010, OttoGermerTheisel2011} have proposed Monte Carlo gradient sampling for visualizing variations of pathlines in 2D and 3D uncertain vector fields. 
 Bhatia et al.~\cite{BhatiaJadhavBremer2012} have studied the  \emph{edge maps} for error analysis of uncertain gradient flows. 
Nagraj et al.~\cite{NagarajNatarajanNanjundiah2011} have proposed a measure to quantify gradient uncertainty for multifield data. 

\para{Uncertainty visualization of topological descriptors.} 
A major challenge in visualizing topological descriptors is to encode data uncertainty. 
Various uncertainty visualization techniques~\cite{Kraus2010, WuZhang2012, ZhangAgarwalMukherjee2015} have been proposed to explore structural variations of contour trees for noisy data. 
Recent work by Lin et al.~\cite{YanWangMunch2019} is the first to study structural averages of merge trees in the context of uncertainty visualization.
The analysis and visualization of topological variations in the context of uncertain data remains an open research challenge~\cite{HeineLeitteHlawitschka2016}. 

Several studies have addressed challenges associated with  level sets visualization in the face of uncertainty, including the \emph{contour boxplots}~\cite{WhitakerMirzargarKirby2013}, probabilistic marching cubes~\cite{PothkowWeberHege2011, PothkowHege2013}, and level set extraction from uncertain data   ~\cite{AthawaleEntezari2013, AthawaleSakhaeeEntezari2016, AthawaleJohnson2019}. 

Multicharts for comparative 3D ensemble visualization~\cite{DemirDickWestermann2014}, dynamic volume lines~\cite{WeissenbockFrohlerGroller2019}, Gaussian mixture data representations \cite{LiuLevineBremer2012}, and statistical volume visualization~\cite{SakhaeeEntezari2017} are a few important contributions in volume rendering for visualizing uncertainty. 

%% file: sec-background.tex
\section{Technical Background}
\label{sec:background}

\para{Morse complexes.}
We focus on the construction of 2D Morse complexes.
For simplicity, let $\Mspace \subset \Rspace^2$ be a 2D smooth manifold with   boundary (we further ignore the boundary condition for most of our discussion). 
Let $f: \Mspace \to \Rspace$ be a Morse function; $\grad{f}$ denotes its gradient. 
A point $x \in \Mspace$ is considered \emph{critical} if $\grad{f}(x)=0$; otherwise it is \emph{regular}.  
At any regular point $x$, the gradient is well defined, and integrating it in both ascending and descending directions traces out an integral line, which is a maximal path whose tangent vectors agree with the gradient \cite{EdelsbrunnerHarerZomorodian2001}.
Each integral line begins and ends at critical points.   
The \emph{descending manifold} surrounding a local maximum is defined as all the points whose integral lines end at the local maximum. 
The descending manifolds decompose the domain into $2$-cells, whereas critical points are the $0$-cells, and integral lines connecting the critical points are tfhe $1$-cells. 
As illustrated in Fig.~\ref{fig:msc}(a), these cells form a complex called a \emph{Morse complex} of $f$, denoted as $\MS = \MS_f$ (whenever $f$ is clear from the context). In particular, all the points inside a single $2$-cell have their local gradient flows (integral lines) ending at the same local maximum. 

Similarly, the \emph{ascending manifold} surrounding a local minimum is defined as all the points whose integral lines begin at the local minimum. 
The ascending manifolds decompose the domain into a dual complex called the Morse complex of $-f$ (Fig.~\ref{fig:msc}(b)), where points in the same $2$-cell have their integral lines originating from the same local minimum.
The set of intersections of ascending and descending manifolds creates the \emph{Morse-Smale} complex of $f$ (Fig.~\ref{fig:msc}(c)) . 

\para{Persistence and persistence simplification.} 
Persistent homology is a tool in topological data analysis for quantifying the significance of topological features. We give a 1D illustrative example of persistence below; see~\cite{EdelsbrunnerHarer2008} for an introduction.

 \begin{figure}[!ht]
   \centering
     \includegraphics[width=0.98\columnwidth]{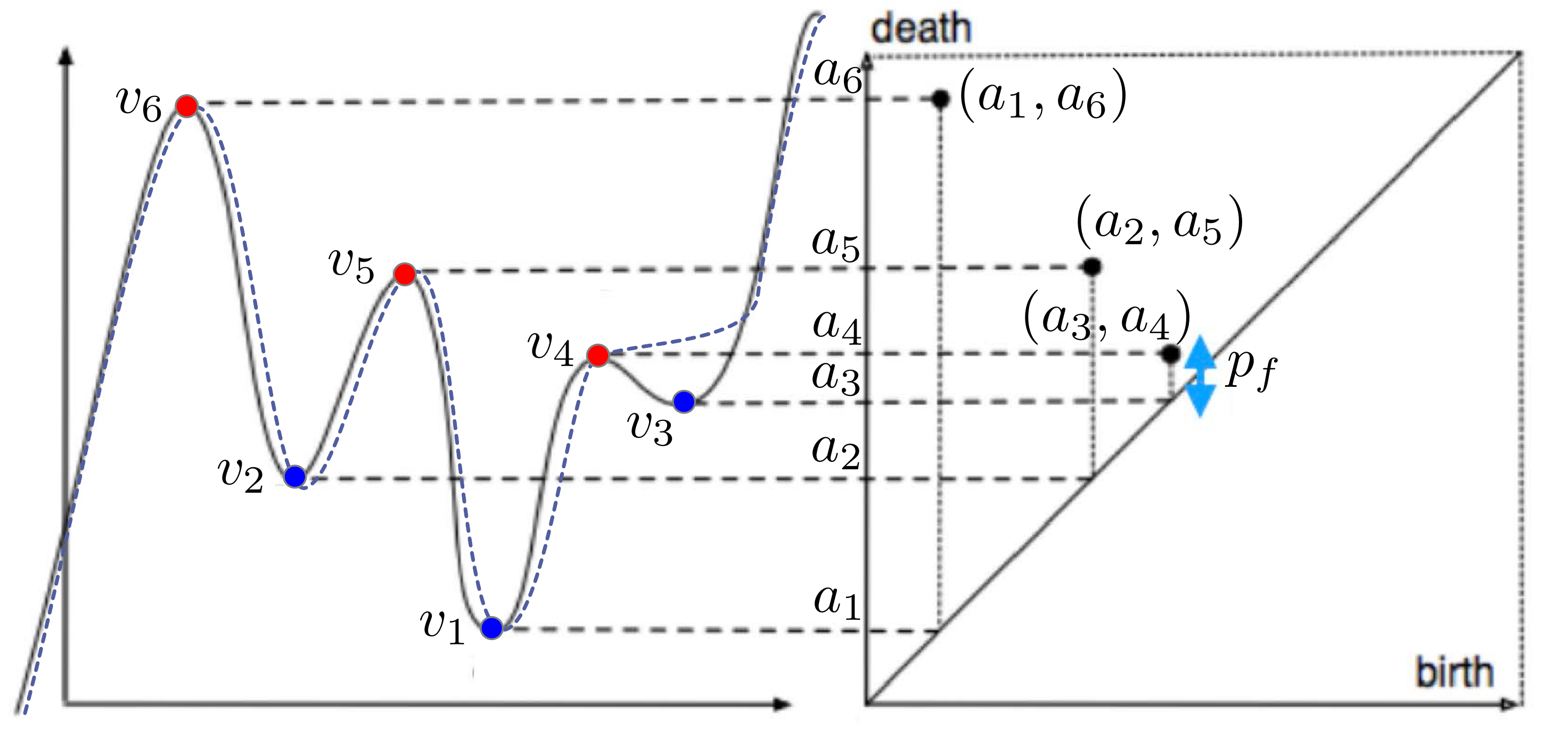}
     \vspace{-2mm}
     \caption{A persistence diagram (right) for a 1D function $f$ (left). $p_f$ denotes the persistence of the smallest topological feature of $f$. Persistence simplification of the smallest topological feature produces a simplified function represented by the (slightly shifted) dotted curve.} 
     \label{fig:persistence}
 \end{figure}

To study the scalar field topology of a 1D Morse function $f: \Mspace \to \Rspace$ (where $\Mspace \subseteq \Rspace$), we focus on the topological structures of its sublevel sets. 
The sublevel sets of $f$ are defined as $\Mspace_a = f^{-1}(\infty, a]$ for any $a \in \Rspace$. 
The $0$D persistent homology tracks the connected components ($0$D topological features) of sublevel sets $\Mspace_a$ as $a$ increases from $-\infty$.
Let $a_1 < a_2 < \cdots < a_m$ be the function values of critical points $v_1, v_2, \cdots, v_m$, respectively.
The collection of sublevel sets forms a filtration $\Mspace_{a_1} \xhookrightarrow{} \Mspace_{a_2} \xhookrightarrow{} \cdots \xhookrightarrow{} \Mspace_{a_m}$ connected by inclusion maps.
Treating $a$ as a \emph{time} parameter, as $a$ increases, components of $\Mspace_a$ appear and disappear  when $a$ passes through critical values. 

As illustrated in Fig.~\ref{fig:persistence}, a 1D function $f$ within its visible domain contains six critical points: local minima are in blue and local maxima are in red. 
As $a$ increases from $-\infty$, a component is born (appears) at time $a_1$ when the sublevel set passes through a local minimum $v_1$. 
Such a component is represented by the critical point $v_1$ that gives birth to it. 
Similarly, a second component is born at $a_2$ and represented by $v_2$; a third component is born at $a_3$ and represented by $v_3$.
At $a_4$, the component represented by $v_3$ and the component represented by $v_1$ merge into one component. 
For consistency, the younger component represented by $v_3$ (the one that is born later) dies (disappears) as a result of the merge. 
In other words, $v_3$ gives rise to a component that is destroyed by $v_4$. 
The critical points $v_3$ and $v_4$ therefore form a persistence pair $(v_3, v_4)$. 
With an abuse of notation, we attach a \emph{persistence measure} to the critical points $v_3$ and $v_3$ that captures the lifespan or significance (called \emph{persistence}) of the topological feature they represent, i.e., their function value difference, $|f(v_4) - f(v_3)| = a_4 - a_3$. 
The birth time and the death time of such a component also give rise to a point $(a_3, a_4)$ in the \emph{persistence diagram} on the plane. 
Similarly, the component that is born at $a_2$ dies at $a_5$, and the component that is born at $a_1$ dies at $a_6$, giving rise to two more points $(a_2, a_5)$ and $(a_1, a_6)$ in the persistence diagram. 

Persistence introduces the notion of scale for learning the structure of a function where small-scale features are commonly considered as noise. 
 Therefore, it is widely used for topological de-noising through persistence simplification~\cite{EdelsbrunnerLetscherZomorodian2002}. 
 As illustrated in Fig.~\ref{fig:persistence}, the critical point pair $(v_3, v_4)$ with the smallest persistence can be simplified producing a simplified function. 

\begin{figure}[!ht]
  \centering
    \includegraphics[width=0.48\textwidth]{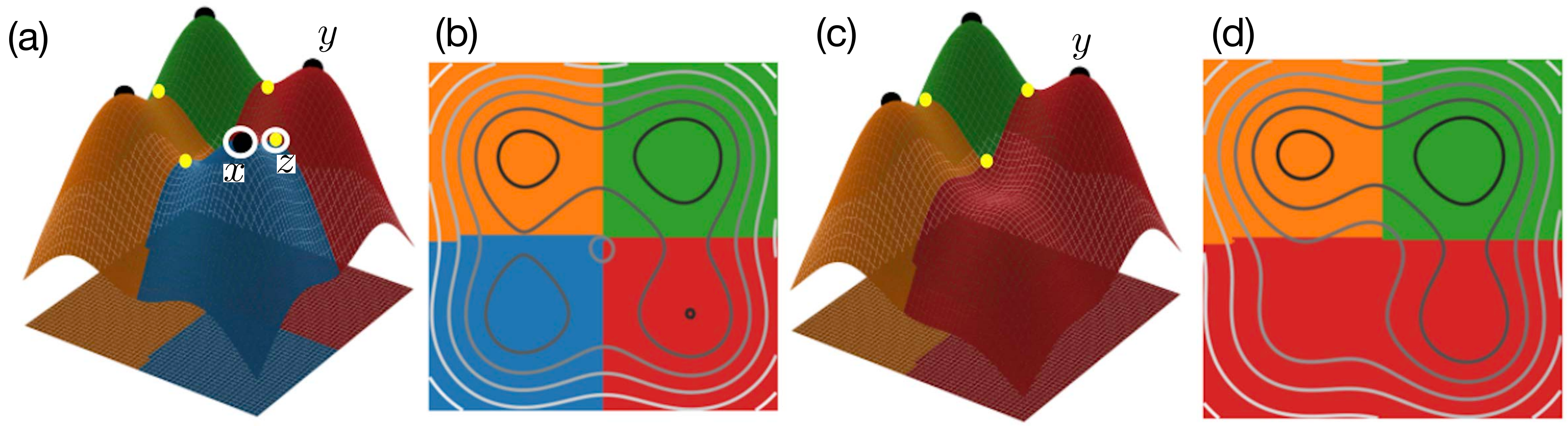}
    \caption{A 2D Morse complex before (a, b) and after (c, d) persistence simplification. Both 3D (a, c) and 2D views (b, d) of the Morse complexes are shown. Black points denote local maxima; yellow points denote local minima or saddles. White circles in (a) represent a saddle-maximum $(z,x)$ pair with the minimal persistence that is simplified in (c).}
    \label{fig:simplification}
\end{figure}

In visualization, persistence has been used to simplify topological structures, such as Morse and Morse-Smale complexes~\cite{EdelsbrunnerHarerZomorodian2003,GyulassyNatarajan2005}. 
The concept of persistence computation in 1D illustrated in Fig. \ref{fig:persistence} can be extended to high-dimensional functions. 
For a 2D scalar function, we create a hierarchical Morse complex~\cite{EdelsbrunnerLetscherZomorodian2002} by simplifying  persistence pairs (in this case, maximum-saddle pairs) in the order of increasing persistence values~\cite{GuntherReininghausSeidel2014}. 
Persistence assigned to each critical point in the complex intuitively describes the scale at which a critical point disappears through simplification.
Persistence pairs can be simplified  by successively canceling pairs of critical points connected in the complex with minimal persistence while avoiding certain degenerate situations (see~\cite{GuntherReininghausSeidel2014} for implementational details). 
 Fig.~\ref{fig:simplification} illustrates the process of persistence simplification for a 2D Morse complex. 
 A saddle-maximum pair $(z,x)$ with the minimal persistence in Fig.~\ref{fig:simplification}a is simplified in Fig.~\ref{fig:simplification}c. 
The simplification merges two 2D Morse complex cells into one; the blue cell represented by local maximum $x$ merges into the red cell represented by local maximum $y$. 
The (ascending) integral lines of all points in the blue cell change their destination from $x$ to $y$.    
We leverage changes in flow directions in the derivation of the {\SM}  (Sec. \ref{sec:survivalMap}).

%% file: sec-methods-pmap.tex
\section{Probabilistic Map} 
\label{sec:probabilisticMap}

For our first approach, we introduce a {\PM} that utilizes positional uncertainties of local maxima across ensemble members to derive structural uncertainty in a Morse complex. 
Following an overview in Sec.~\ref{sec:introduction}, we describe our pipeline via a synthetic dataset, called the Ackley dataset. 

\begin{figure}[!ht]
\centering 
\includegraphics[width=0.8\columnwidth]{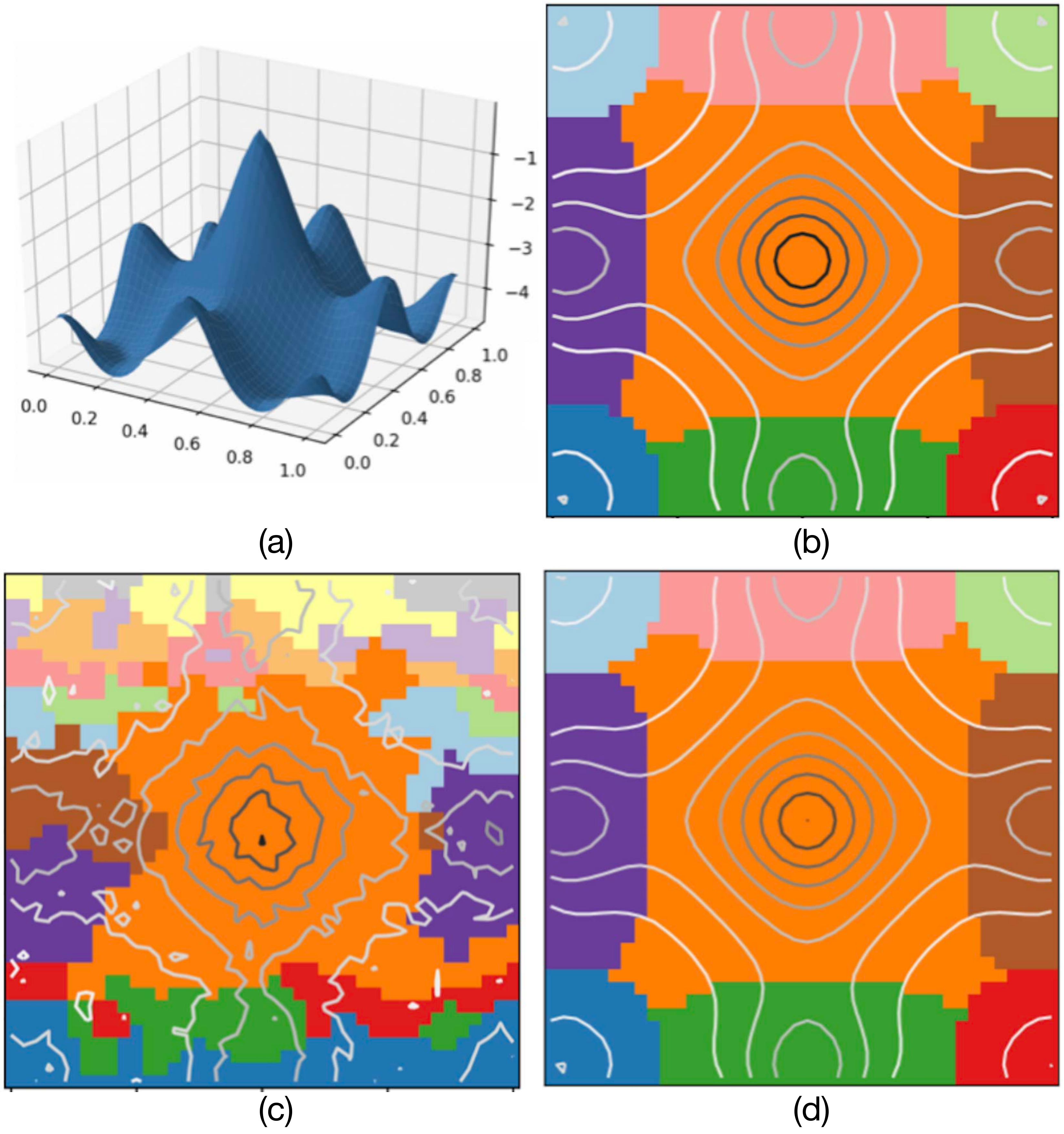}
\vspace{-4mm}
\caption{Morse complexes of the Ackley dataset. (a) A 3D visualization of the ground truth Ackley function $f$. (b) The Morse complex of $f$ with nine $2$D cells surrounding local maxima. (c) A Morse complex generated by mixing $f$ with uniform noise. (d) The Morse complex of the mean field from an ensemble of scalar fields generated as in (c). Contours are visualized in various shades of grey.}
\label{fig:AckleyMorseComplex}
\end{figure}

Fig.~\ref{fig:AckleyMorseComplex}a visualizes the Ackley function~\cite{Ackley1987} $f$ as the ground truth. 
$f$ is made into a Morse function using simulation of simplicity~\cite{EdelsbrunnerMucke1990}.
$f$ contains nine local maxima, which produces nine 2-cells in its corresponding Morse complex in Fig.~\ref{fig:AckleyMorseComplex}b.  
Let $p_f$ be the persistence of the smallest ($0$-cells) topological feature in $f$.
We generate an ensemble of uncertain scalar fields $\{f_i\}_{i=1}^{n}$ by mixing $f$ with  noise sampled from a uniform distribution $\epsilon_i(x) \sim U(0, 0.6 \times p_f/2)$; an ensemble member is shown in Fig.~\ref{fig:AckleyMorseComplex}c. 
For comparison, we compute the mean field of the ensemble, $\bar{f}=\frac{1}{n}\sum_{i}f_i$ and visualize its Morse complex in Fig.~\ref{fig:AckleyMorseComplex}d. The Morse complex of $\bar{f}$ appears similar to the ground truth; however, it does not capture structural variations on the boundaries of Morse complex cells. 

\begin{figure}[!h]
\centering 
\includegraphics[width=0.98\columnwidth]{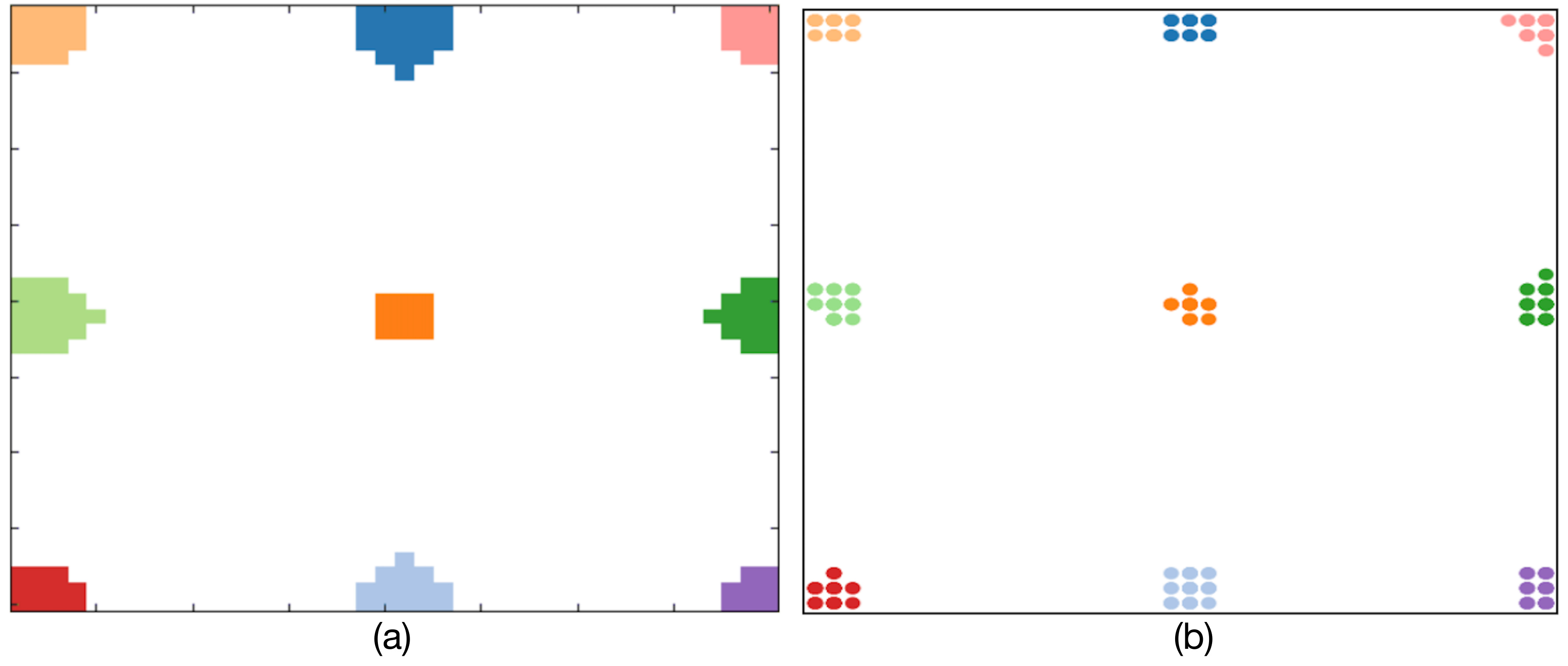}
\vspace{-2mm}
\caption{(a) Nine mandatory maxima of an ensemble of uncertain Ackley functions. (b) A clustering of local maxima into nine clusters across ensemble members after persistence simplification.}
\label{fig:ackleyMCP}
\end{figure}

\para{Computing the \PM.} First, we compute mandatory maxima for an ensemble of uncertain Ackley functions $\{f_i\}_{i=1}^{n}$, resulting in $l = 9$ mandatory local maxima; see Fig.~\ref{fig:ackleyMCP}a. 
Under our noise model (see Sec.~\ref{sec:noiseModel} for details), the number of mandatory maxima $l$ is consistent with the number of local maxima in the ground truth function $f$. 
We assign a label to each mandatory local maxima (and equivalently, to each local maxima of $f$); let $[l] = \{1, 2, \cdots, l\}$ denote the set of labels. In Fig.~\ref{fig:ackleyMCP}a, the labels of mandatory maxima are represented by different colors.

Second, we apply persistence simplification to the Morse complex~\cite{GuntherReininghausSeidel2014} of each $f_i$ until we are left with $l$ local maxima for each ensemble member.
We then cluster $n \times l$ local maxima (obtained after simplification) across all ensemble members into $l$ clusters; see Fig.~\ref{fig:ackleyMCP}b. 

Third, for each point in the domain, we compute a probability distribution of its cluster membership across the ensemble. 
That is, fix an ensemble member $\MS_i$ that arises from $f_i$, we trace the ascending integral line of each point $x \in \Mspace$ toward its destination, a local maximum $y \in \Mspace$, and assign to $x$ the label of $y$ as its cluster membership; let $\alpha_i: \Mspace \to [l]$ denote such an assignment. 

\begin{figure}[!ht]
  \centering
    \includegraphics[width=0.48\textwidth]{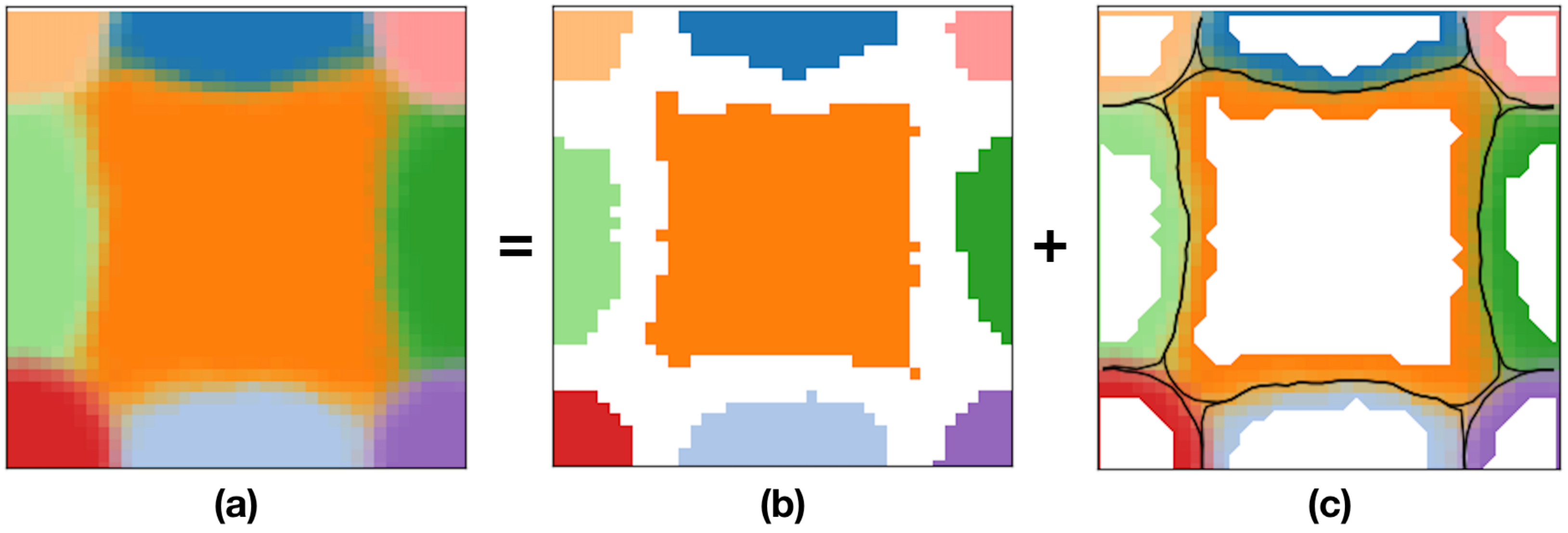}
    \vspace{-2mm}
    \caption{(a) The visualization of {\PM} $\PMap$ for the Ackley dataset. 
    (b) Colored regions contain points with certainty in the domain; white are points with uncertainty.
    (c) Points with uncertainty are further visualized based on their proximity to the points with certainty through color blending.}
    \label{fig:approach2}
\end{figure}

Finally, the {\PM} $\PMap: \Mspace \to \Rspace^{l}$ is defined to be a discrete probability distribution of label assignments for each $x \in \Mspace$.  
Each $x \in \Mspace$ is assigned $n$ labels $\{\alpha_1(x), \cdots, \alpha_n(x)\}$ across ensemble members.  
Let $\PMap_j(x)$ be the number of times $x$ is assigned a label $j \in [l]$ divided by $n$. 
Then $\PMap(x) = (\PMap_1(x), \cdots, \PMap_l(x))$. 
For a point $x \in \Mspace$, if $\PMap_j(x)=1$ (implying $\PMap_k = 0$ for all $k \neq j$) for some $j$, then $x$ is a \emph{point with certainty}; otherwise it is a \emph{point with uncertainty}.  
Points with certainty are those whose gradient flows to the same mandatory local maximum across ensemble members.
Points with uncertainty are those whose flow behaviors vary across ensemble members. 

\para{Visualizing the \PM.} Fig.~\ref{fig:approach2}a visualizes $\PMap$ for the Ackley dataset. 
Fig. \ref{fig:approach2}b shows the points with certainty in color; for example, all points in the orange region have their gradients flow to the same mandatory local maximum. 
The white regions are points with uncertainty. 
Points with uncertainty are further visualized in Fig.~\ref{fig:approach2}c based on their proximity to the points with certainty; for example, the orange points in Fig.~\ref{fig:approach2}c are the points that have {higher probabilities of flowing to the orange cluster shown in Fig.~\ref{fig:ackleyMCP}b than the other nearby clusters. 
For a pair of adjacent regions with different labels $i$ and $j$ (e.g., orange vs. light green), a black contour contains all points $x \in \Mspace$ such that $P_i(x) = 0.5$ for some cluster label $i$; we refer to such black contours as the \emph{expected boundaries}.

We employ color blending to visualize $\PMap$.  
Let $\CMap: \Mspace \to \Rspace$ be the coloring function.  
Suppose each mandatory local maximum is assigned a color, $\{c_1, \cdots, c_l\}$, where $c_i \in \Rspace$. 
$\PMap_i(x)$ is the probability of a point $x \in \Mspace$ having its gradient flow terminate in a mandatory local maximum with the label $i$.
Point $x$ is then assigned a color $\CMap(x) = \Sigma_{i} c_i \PMap_i(x)$. 

\begin{figure}[!ht]
  \centering
    \includegraphics[width=0.48\textwidth]{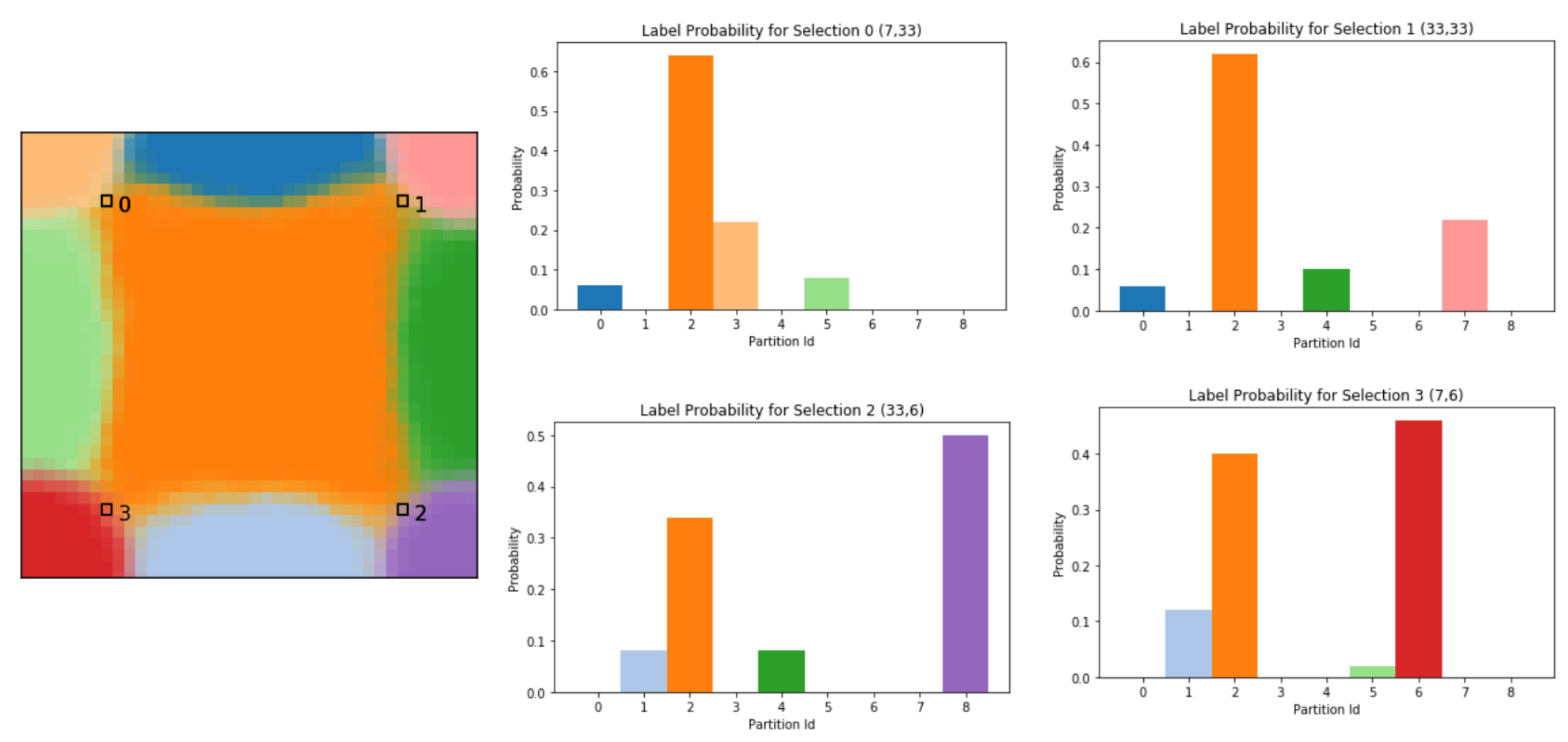}
    \vspace{-2mm}
    \caption{Interactive queries for the {\PM} of the Ackely dataset. 
    Four query locations labeled $0$ to $3$ in the domain are selected.
    The discrete probability distribution $\PMap$ associated with each query location is visualized using a bar chart. \emph{Partition Id} corresponds to the label of each mandatory local maximum.}
    \label{fig:pdfQueryAckley}
\end{figure}

\para{Interactive Queries.}
To further understand the points with uncertainty in $\PMap$ (Fig.~\ref{fig:approach2}a), we provide interactive queries based on the framework of Potter et al.~\cite{PotterKirbyXiu2012}, which focuses on interactive visualization of probability and cumulative density functions. 
Fig.~\ref{fig:pdfQueryAckley} illustrates such interactive queries for a {\PM} of the Ackley dataset for four point locations with uncertainty. 
As illustrated by the bar charts, both points $0$ and $1$ are dominantly  orange, while point $2$ is mainly violet and orange, and point $3$ is  primarily red and orange.  
Interactive queries not only shed light on flow uncertainties but also enable us to adjust cluster membership for an ensemble.

\begin{figure}[!ht]
  \centering
    \includegraphics[width=0.4\textwidth]{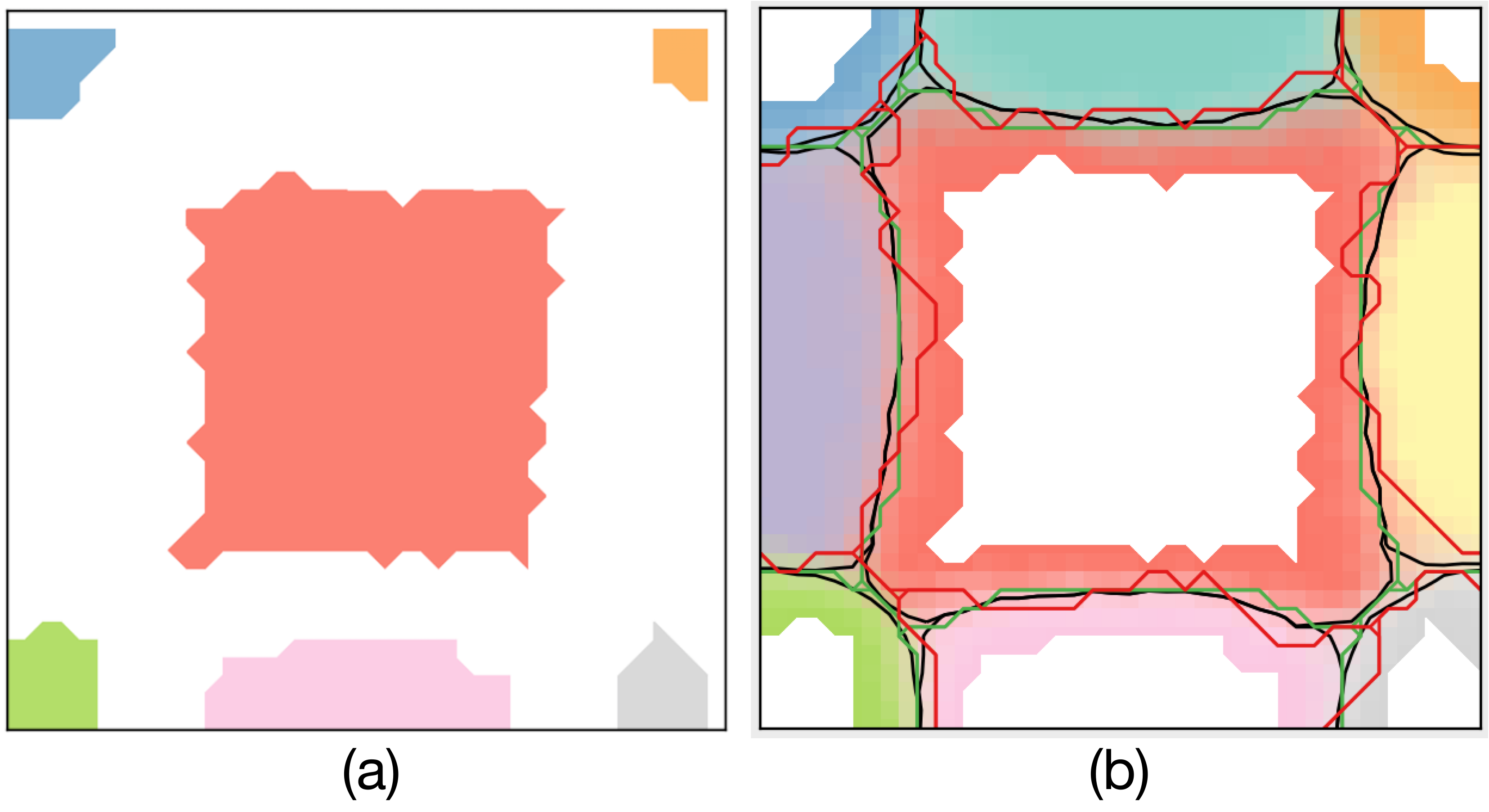}
    \vspace{-4mm}
    \caption{The {\PM} visualization of the Ackley dataset for nonparametric noise distribution. (a) Colored regions contain points with certainty. (b) Colored regions contain points with uncertainty. Green boundaries are from the ground truth function; red boundaries are from the mean field; and black boundaries are the expected boundaries, that is, those points with $\PMap_i = \sum_{j\neq i}\PMap_j = 0.5$ for some $i$.}
    \label{fig:meanVsExpected}
\end{figure}

\para{{\PM} for Nonparametric Distributions.}
We compare the {\PM} vs. the Morse complex of the mean field in Fig.~\ref{fig:meanVsExpected} for nonparametric, multimodal noise distributions. 
The regions with uncertainty in $\PMap$ are overlaid with the Morse complex boundaries computed from the ground truth (green), the mean field (red), and the points with $\PMap_i = \sum_{j\neq i}\PMap_j = 0.5$ (black). 
The black (expected) boundaries from the {\PM} lie closer to the ground truth than the red boundaries from the mean field. 
The {\PM}, therefore, not only captures positional variations but also provides reasonable approximations of Morse complex cell boundaries.

%% file: sec-methods-smap.tex
\section{Survival Map}
\label{sec:survivalMap}

For our second approach, we introduce a {\SM} to quantify structural deviations in local gradient flows across ensemble members (see Fig.~\ref{fig:pipeline}d). 
Instead of studying the spatial correlations among local maxima, we study directional changes of gradient flows as a result of persistence simplification~\cite{EdelsbrunnerLetscherZomorodian2002, GuntherReininghausSeidel2014}.

For each ensemble member $\MS_i$ that arises from a function $f_i$, we apply a hierarchical persistence simplification (Fig.~\ref{fig:pipeline}d) of $\MS_i$.  
In the case of a Morse complex, we focus on canceling maximum-saddle pairs until only the global maximum remains. 
We introduce a \emph{survival measure} for each point $x \in \Mspace$ based on how frequently it changes its local gradient flows during the  simplification process; let $\beta_i: \Mspace \to \Rspace$ denote such an assignment. 
Let $\{\lambda_1, \cdots, \lambda_{n_i}\}$ denote the persistence of maximum-saddle pairs to be cancelled in increasing order, where $n_i+1$ is the total number of local maxima in $f_i$.
We now describe our algorithm in detail. 

First, we compute $\beta_i: \Mspace \to \Rspace$ for each ensemble member. 
We initialize $\beta_i$ to be zero everywhere. 
We perform $n_i$ steps of persistence simplification. 
For each step $j$ ($1 \leq j \leq n_i$), we cancel the maximum-saddle pair $(v_j, u_j)$ with the lowest persistence value $p = \lambda_j$ (see  Fig.~\ref{fig:simplification}). 
As a result, the gradient flows surrounding the local maximum $v_j$ are redirected to a nearby local maxima $v_k$, effectively merging the $2$-cell sounding $v_j$ into the $2$-cell surrounding $v_k$. 
For all points $x$ in the $2$-cell surrounding $v_k$, we increase $\beta_i(x)$ by $\lambda_j$. 
Intuitively, $\beta_i$ is incremented within a local neighborhood of $v_k$ where the gradient flow directions \emph{survive} (remain unchanged) after persistence simplification.   
The above process is repeated until $j = n_i$, i.e., when the entire Morse complex is simplified into a single $2$-cell surrounding the global maximum. 

Fig. \ref{fig:simplification} illustrates one step of our algorithm with a toy example. 
Suppose we simplify the maximum-saddle pair $(x, z)$ with the lowest  persistence $\lambda$; this means that the blue 2-cell merges into the red 2-cell after persistence simplification. 
$\beta: \Mspace \to \Rspace$ is increased by $\lambda$ for all points in the red 2-cell of Fig. \ref{fig:simplification}a, and it is unchanged everywhere else in the domain. 
$\beta$ therefore captures the \emph{survivability} of local gradient flows after persistence simplification. 

\begin{figure}[!h]
\centering 
\includegraphics[width=0.90\columnwidth]{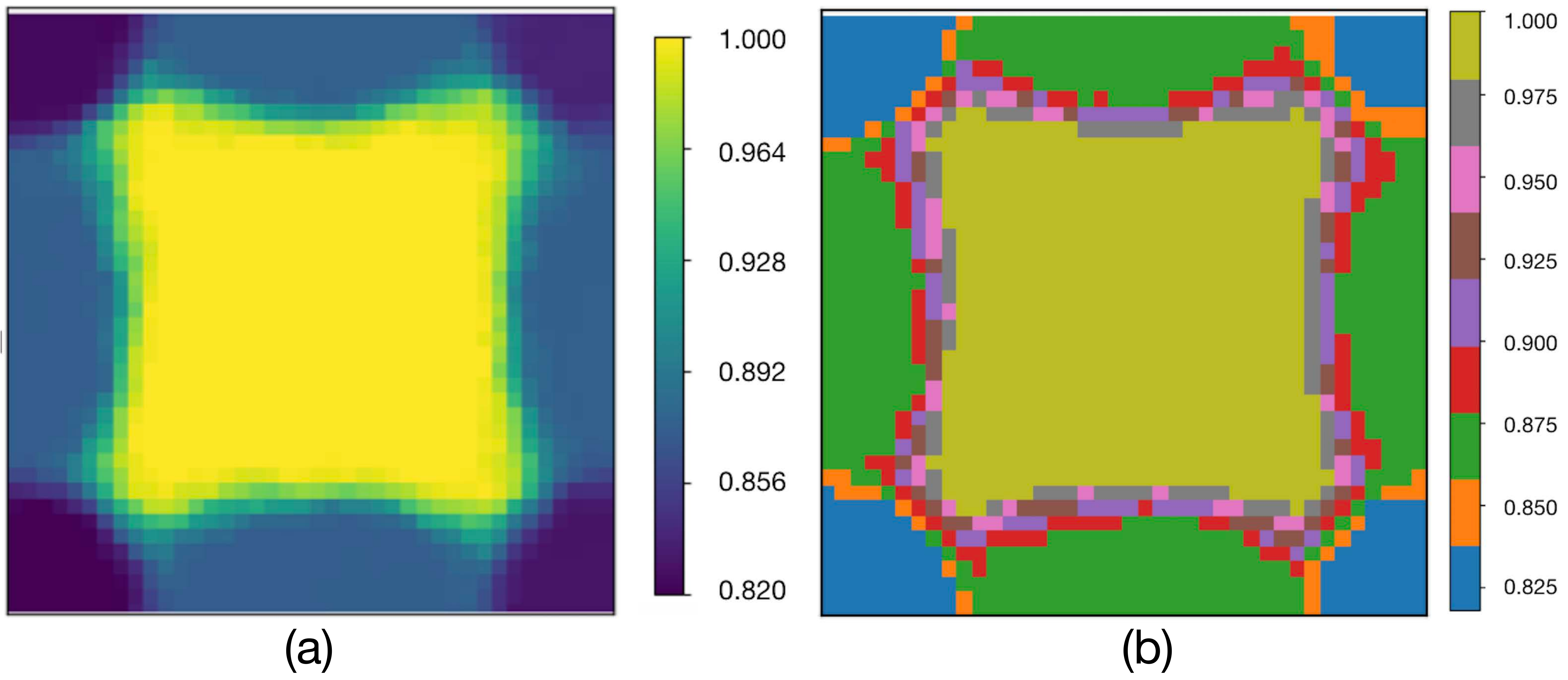}
\vspace{-2mm}
\caption{The visualization of the {\SM} $\SMap$ for the Ackley dataset. (a) $\SMap$ is visualized using a heat color map: yellow means high and blue means low survival values. (b) A quantized visualization of $\SMap$ using nine intervals.}
\label{fig:survivalMapAckley}
\end{figure}

Second, the {\SM} $\SMap: \Mspace \to \Rspace$ is computed as the average value of survival measures across the ensemble $\{\beta_1(x), \beta_2(x), \cdots, \beta_n(x)\}$ for each $x \in \Mspace$, that is, $\SMap(x) = \frac{1}{n}\sum_{i=1}^{n} \beta_i(x)$.
Fig.~\ref{fig:survivalMapAckley} shows the {\SM} of the Ackley dataset using heat color map. 
The yellow region suggests the existence of a relatively tall peak, and the dark blue regions represent the existence of relatively low peaks across all ensemble members. 
This behavior is consistent with the ground truth Ackley function depicted in Fig.~\ref{fig:AckleyMorseComplex}a.

\para{Quantized visualization.}
To gain insight into the variability of $1$-cells as boundaries of the $2$-cells in the {\SM}, we employ a quantized visualization to further differentiate the regions with uncertainty. 
In particular, we divide the range of $\SMap$ into a fixed number of intervals and visualize the pre-image of each interval using a miscellaneous color map.
For example, as shown in Fig.~\ref{fig:survivalMapAckley}(b), the regions with higher color fluctuations indicate positions with higher uncertainty in gradient flow directions.

%% file: sec-results.tex
\section{Results}
\label{sec:results}

We demonstrate the utility of our proposed statistical summary maps for gaining insights into Morse complex uncertainty for synthetic and real-world datasets. 

\subsection{Himmelblau's Function Dataset}
\begin{figure}[!h]
\centering 
\includegraphics[width=0.98\columnwidth]{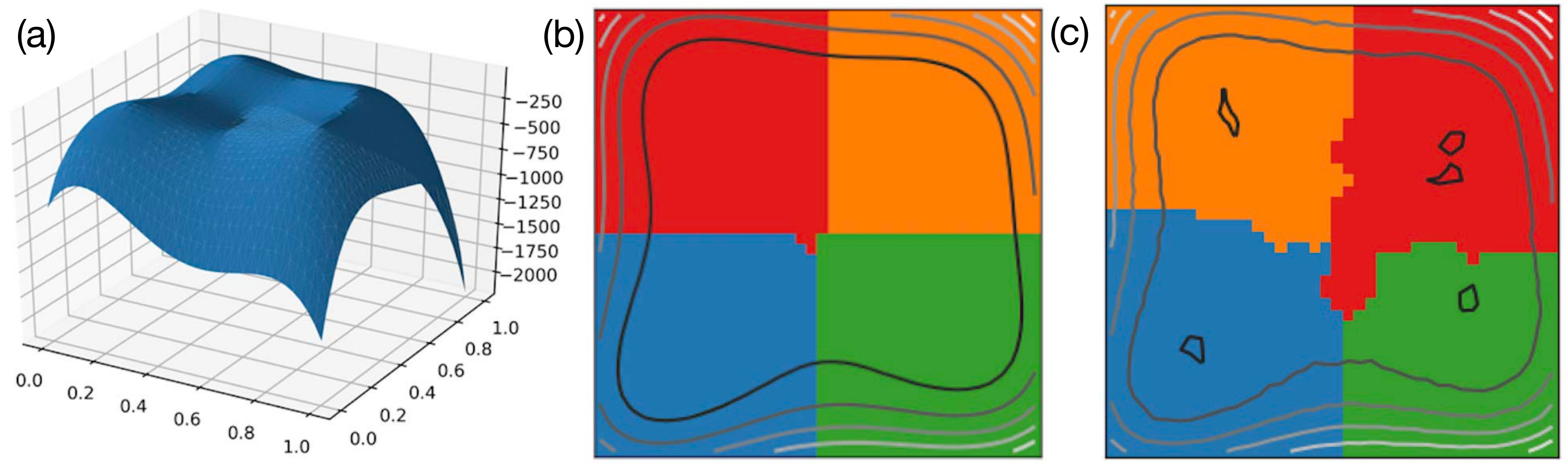}
\vspace{-2mm}
\caption{Morse complexes of the Himmelblau's function dataset. (a) A 3D visualization of the ground truth Himmelblau' function $f$. (b) The Morse complex of $f$ with four $2$-cells. (c) The Morse complex of the mean field from an ensemble of scalar fields generated from $f$ with noise.}
\label{fig:himmelblauPersistenceSurvival}
\end{figure}

Fig.~\ref{fig:himmelblauPersistenceSurvival} visualizes the Morse complexes for the Himmelblau's function dataset~\cite{Himmelblau1972}. 
The Himmelblau's function is a multi-modal function containing four local maxima with equal heights; see Fig.~\ref{fig:himmelblauPersistenceSurvival}a; for our purpose, it is perturbed to be a Morse function using simulation of simplicity~\cite{EdelsbrunnerMucke1990}. 
We generate an ensemble of functions from $f$ with noise $\epsilon < p_f/2$, and visualize the Morse complex of the mean field (Fig.~\ref{fig:himmelblauPersistenceSurvival}c). 
The $2$-cells in the Morse complex of the mean field have distorted boundaries in comparison with the ground truth (Fig.~\ref{fig:himmelblauPersistenceSurvival}b),  and they do not give any insight into the positional uncertainties of such boundaries. 

\begin{figure}[!ht]
  \centering
    \includegraphics[width=0.48\textwidth]{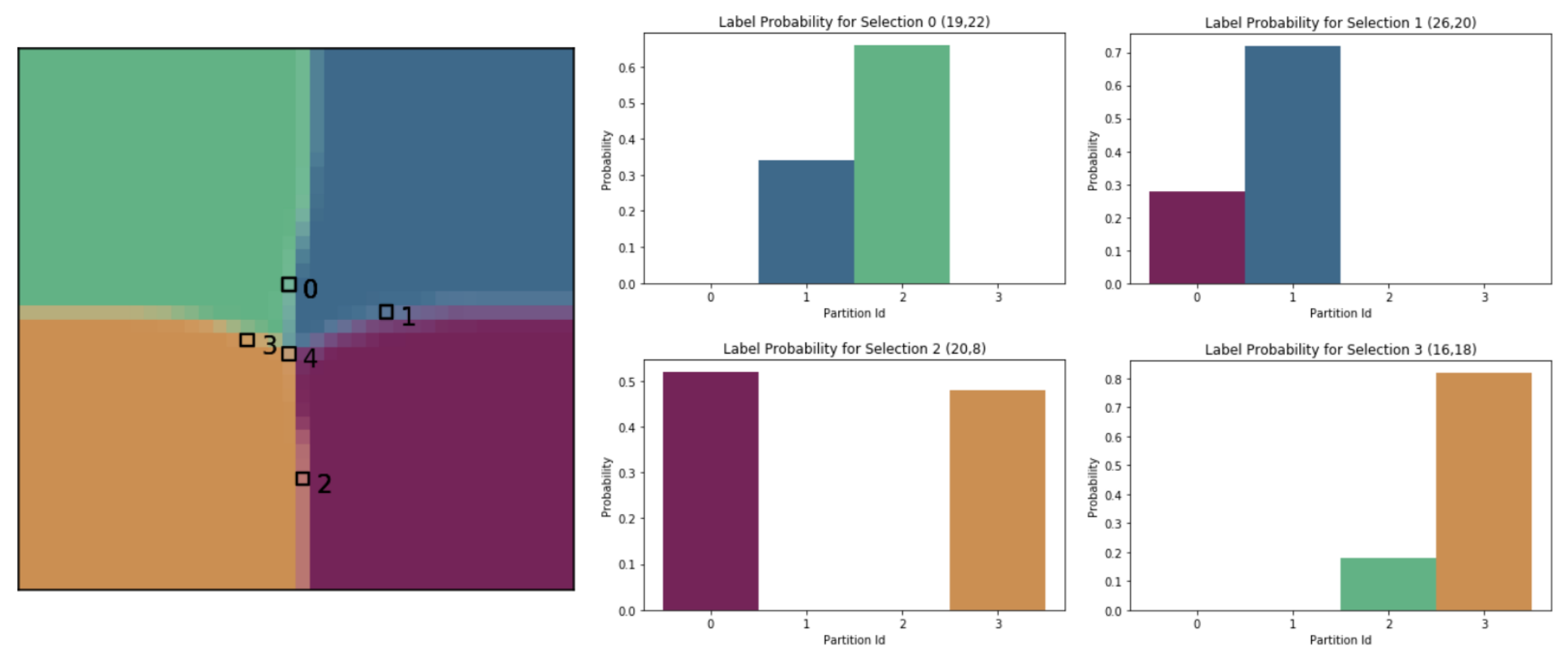}
    \vspace{-4mm}
    \caption{Interactive queries for the {\PM} of Himmelblau's function dataset.}
    \label{fig:pdfQueryHimmelblau}
\end{figure}

\begin{figure}[!ht]
\centering 
\includegraphics[width=0.98\columnwidth]{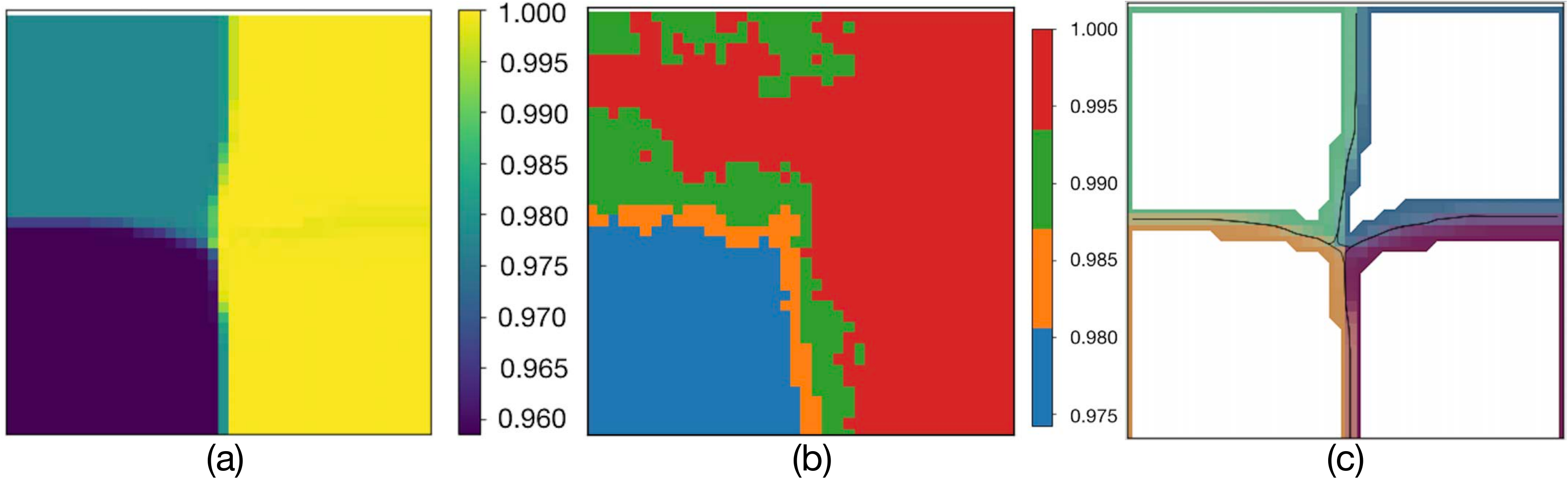}
\vspace{-2mm}
\caption{(a) The visualization of the {\SM} $\SMap$ for the Himmelblau's function dataset. (b) A quantized visualization of  $\SMap$ using four intervals. (c) Colored regions contain points with uncertainty for the {\PM} visualized in Fig.\ref{fig:pdfQueryHimmelblau}. Black contours represent expected boundaries with $\PMap_i = 0.5$ for some cluster $i$.}
\label{fig:thresholdedHimmelblau}
\end{figure}

The {\PM} $\PMap$ and interactive queries at four locations are illustrated in Fig.~\ref{fig:pdfQueryHimmelblau}. The {\SM} $\SMap$ visualized in Fig.~\ref{fig:thresholdedHimmelblau}a has three colored regions. 
The orange and green regions in the ground truth (Fig.~\ref{fig:himmelblauPersistenceSurvival}b) are perceived as a single yellow region in Fig.~\ref{fig:thresholdedHimmelblau}a, which contains local maxima with similar heights across ensemble  members. 
The fuzzy regions in Fig.~\ref{fig:thresholdedHimmelblau}a give insight into the flow uncertainty within the ensemble. 
A quantized visualization of $\SMap$ in Fig.~\ref{fig:thresholdedHimmelblau}b further highlights the flow variations in regions with uncertainty. 
Note that the survival measures attain a narrower range (from $0.96$ to $1$) for the Himmelblau's functions compared to the ones for the Ackely function (Fig.~\ref{fig:survivalMapAckley}a). 
In Fig.~\ref{fig:thresholdedHimmelblau}c, the black contours represent the expected boundaries for the {\PM} visualized in Fig.~\ref{fig:pdfQueryHimmelblau}.

\subsection{K\'{a}rm\'{a}n Vortex Street Dataset}\label{sec:karmanVortexStreet}
\begin{figure}[!ht]
  \centering
    \includegraphics[width=0.49\textwidth]{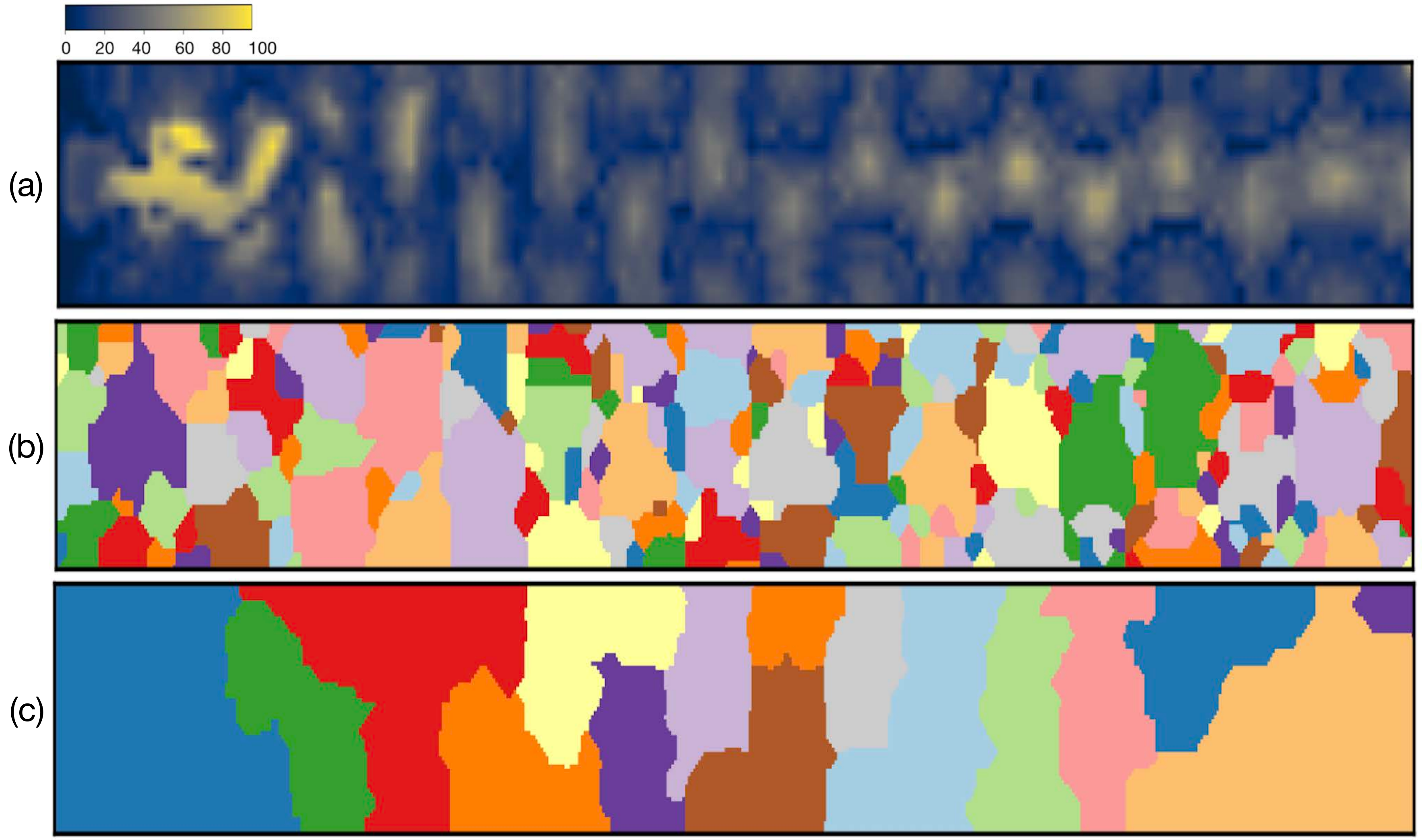}
    \vspace{-6mm}
    \caption{Visualization of the mean field (a) of the K\'{a}rm\'{a}n Vortex Street ensemble dataset together with the Morse complexes of the mean field before (b) and after (c) persistence simplification.}
    \label{fig:flowHeatMap}
\end{figure}

In our first real-world example, we work with the K\'{a}rm\'{a}n Vortex Street ensemble dataset. 
The original flow simulation of K\'{a}rm\'{a}n Vortex Street is available via the Gerris software~\cite{Popine2003}; it is generated as a result of a steady flow (moving from left to right) obstructed by an obstacle situated at the far left. 
We generate an ensemble of scalar fields representing uncertainty in flow velocity by perturbing the uncertain fluid viscosity parameter. 
Each ensemble member is computed as the magnitude of the flow velocity after perturbation. 
We first compute the mean field of the ensemble, as illustrated in Fig.~\ref{fig:flowHeatMap}a. 
Yellow regions indicate vortical structures in the flow. 
The Morse complexes of the mean field before and after persistence simplification are shown in Fig.~\ref{fig:flowHeatMap}b and Fig.~\ref{fig:flowHeatMap}c, respectively. 

\begin{figure}[!ht]
  \centering
    \includegraphics[width=0.49\textwidth]{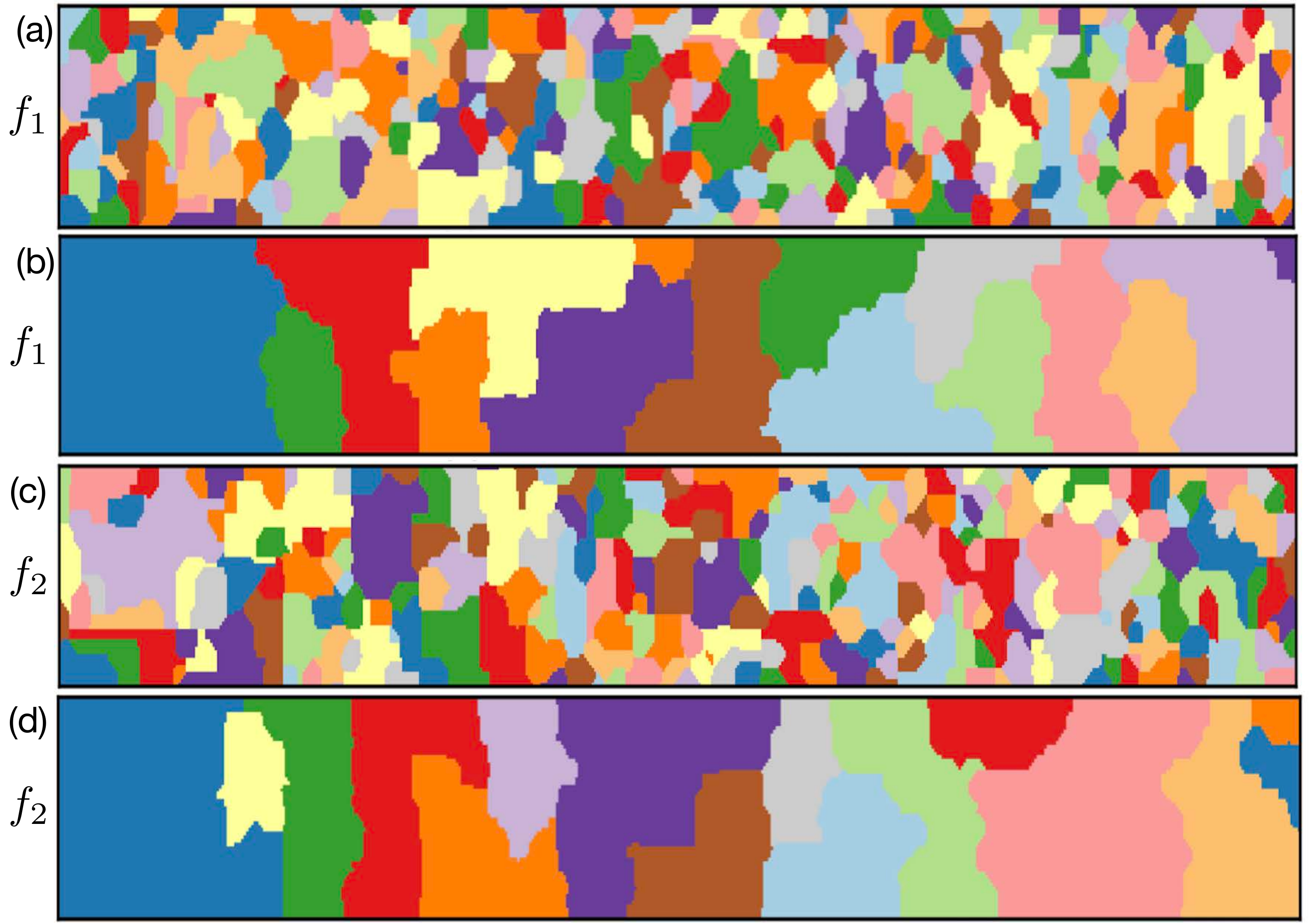}
    \vspace{-4mm}
    \caption{Morse complexes of two ensemble members $f_1$ and $f_2$, before (a, c) and after (b, d) persistence simplification.}
    \label{fig:flowRealizations}
\end{figure}

Even though persistence simplification (Fig.~\ref{fig:flowHeatMap}c) of the mean field gives rise to fewer $2$-cells and a high-level view of its gradient behavior, it does not capture positional uncertainties of the 2-cell boundaries. 
Fig.~\ref{fig:flowRealizations} visualizes Morse complexes of two ensemble members. The positional variations of 2-cell boundaries appear to be substantial, even after persistence simplification. 

\begin{figure}[!h]
  \centering
    \includegraphics[width=0.49\textwidth]{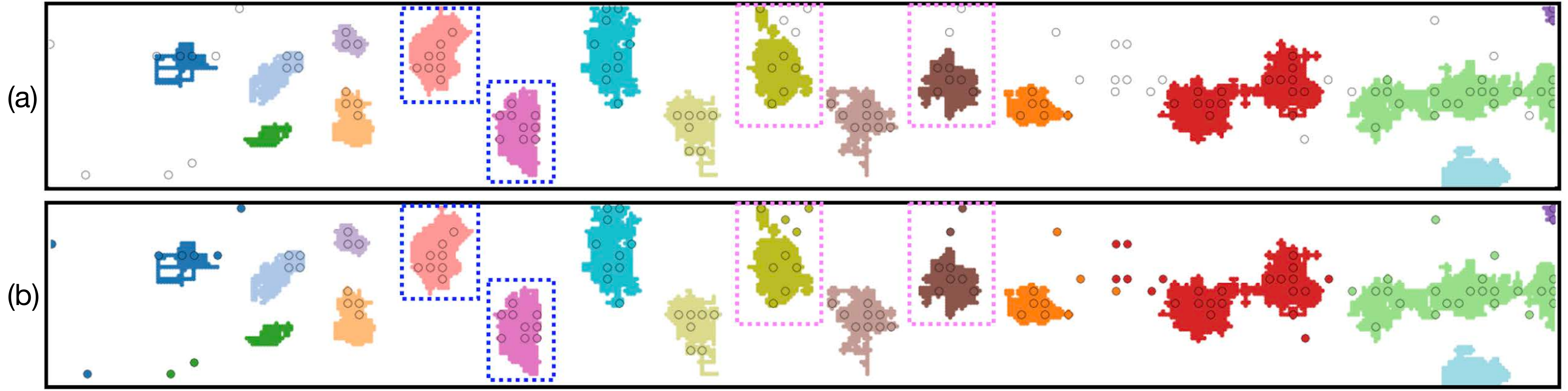}
    \caption{(a) Mandatory local maxima (colored regions) of the K\'{a}rm\'{a}n Vortex Street ensemble dataset. Hollow circles represent local maxima of each ensemble member after simplification. (b) Each circle is assigned the label of its nearest mandatory local maximum.
    }
    \label{fig:mcpFlow}
\end{figure}

We first compute the {\PM} $\PMap$ for the K\'{a}rm\'{a}n Vortex Street ensemble dataset. Fig.~\ref{fig:mcpFlow}a visualizes mandatory local maxima of the ensemble, which form $16$ clusters. Each cluster is assigned a unique color. 
Based on our noise model, we simplify the Morse complex for each ensemble member until $16$ local maxima are left. 
For each ensemble member after simplification, we overlay its local maxima (hollow circles) with the mandatory local maxima (colored regions) in Fig. \ref{fig:mcpFlow}a. 
The dotted blue boxes enclose locations where local maxima for all ensemble members are contained within mandatory maxima clusters; whereas the dotted pink boxes enclose locations where local maxima for a few ensemble members are not contained in any mandatory maxima clusters (see Sec.~\ref{sec:noiseModel} for details). We assign labels (colors) to the (hollow) local maxima based on the labeling of their nearest mandatory local maxima in Fig. \ref{fig:mcpFlow}b.

\begin{figure}[!ht]
  \centering
    \includegraphics[width=0.49\textwidth]{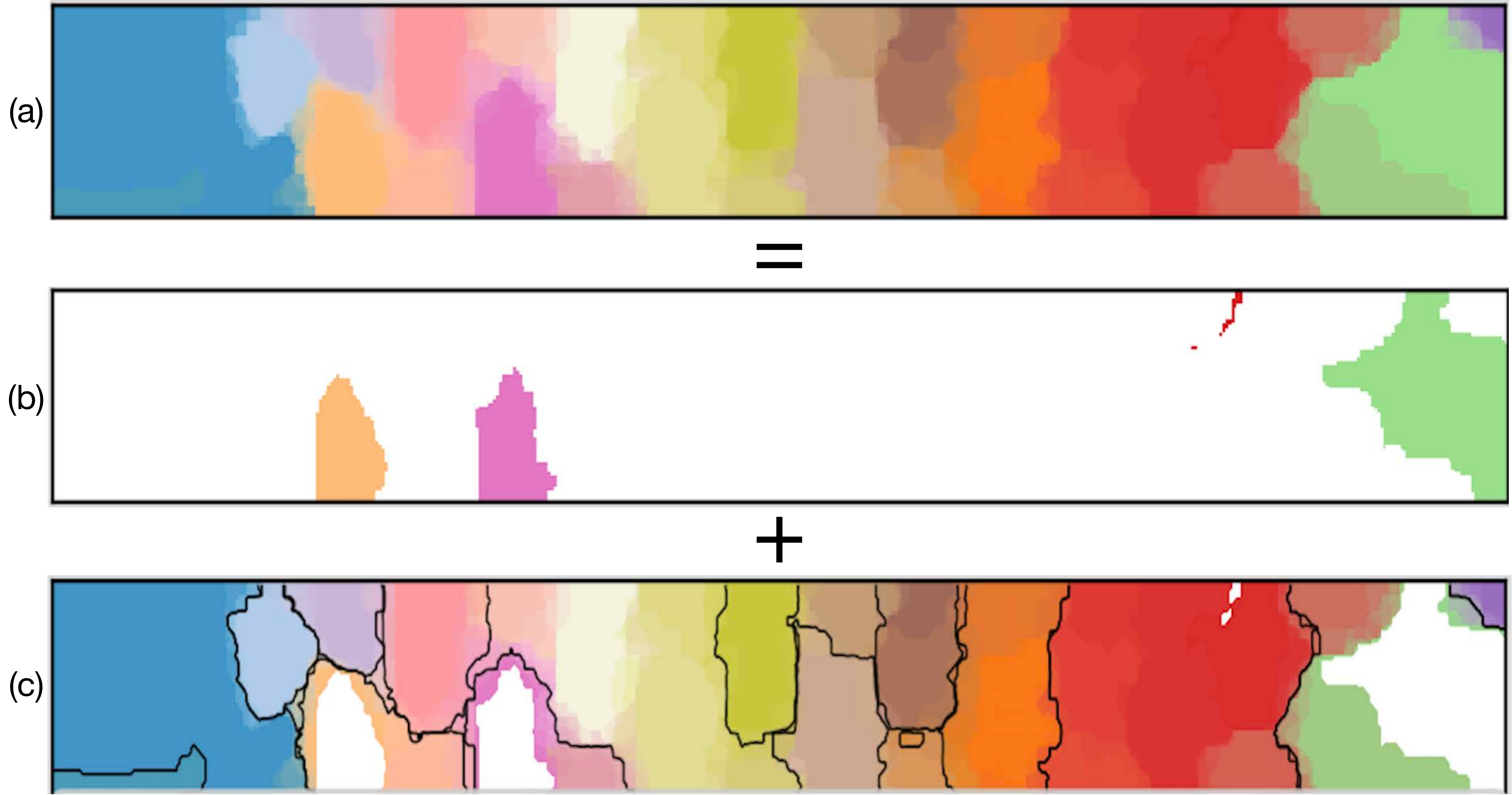}
    \caption{(a) Visualization of the {\PM} $\PMap$ for the K\'{a}rm\'{a}n Vortex Street ensemble dataset. (b) Colored regions contain points with certainty in the domain; white regions include points with uncertainty. (c) Points with uncertainty are further visualized through color blending.
}
    \label{fig:approach2Flow}
\end{figure}

\begin{figure}[!ht]
  \centering
    \includegraphics[width=0.49\textwidth]{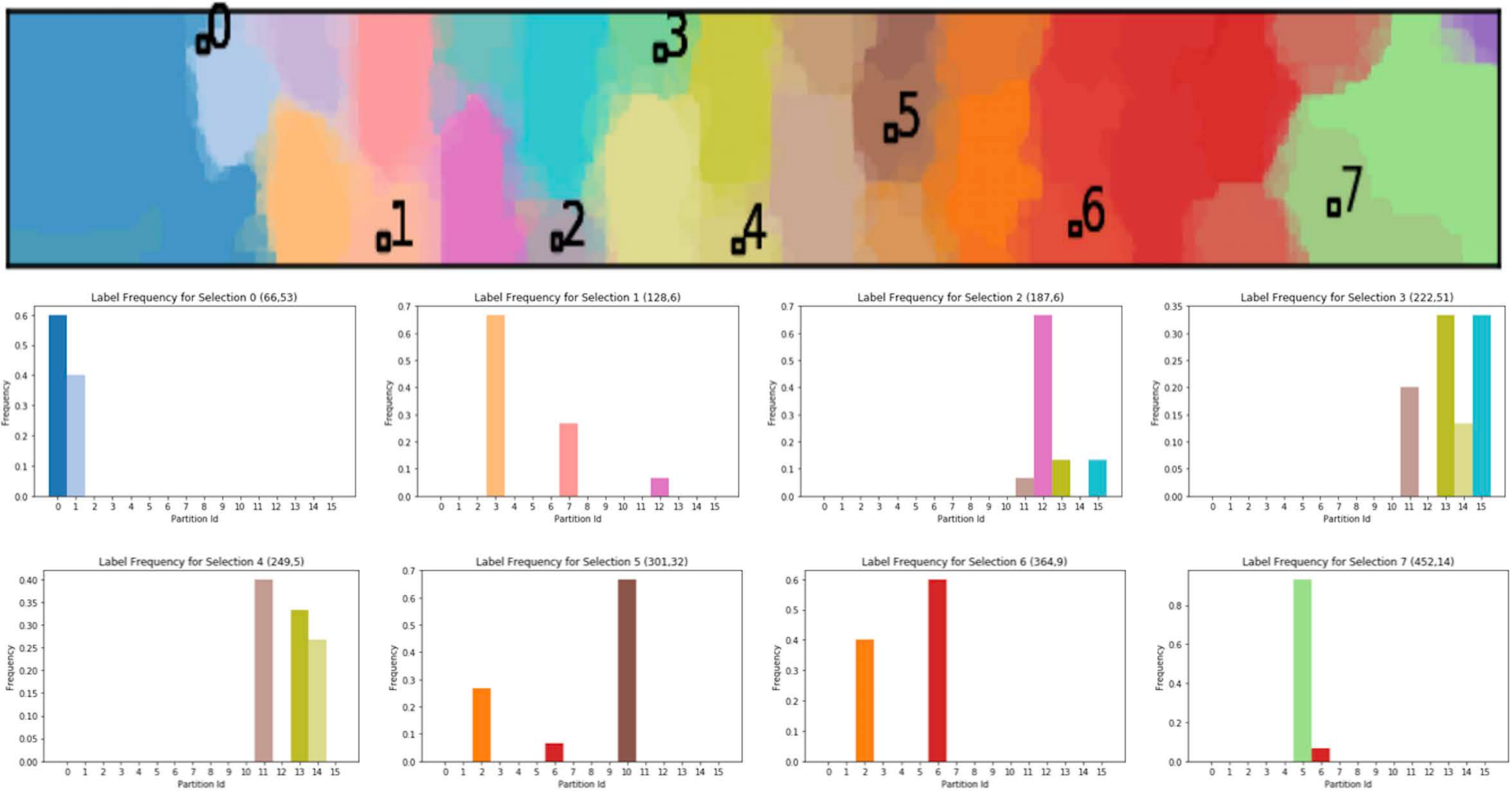}
    \caption{Interactive queries for the uncertain regions of the {\PM} for the K\'{a}rm\'{a}n Vortex Street ensemble dataset.}
    \label{fig:pdfQueryFlow1}
\end{figure}

\begin{figure}[!ht]
  \centering
    \includegraphics[width=0.47\textwidth]{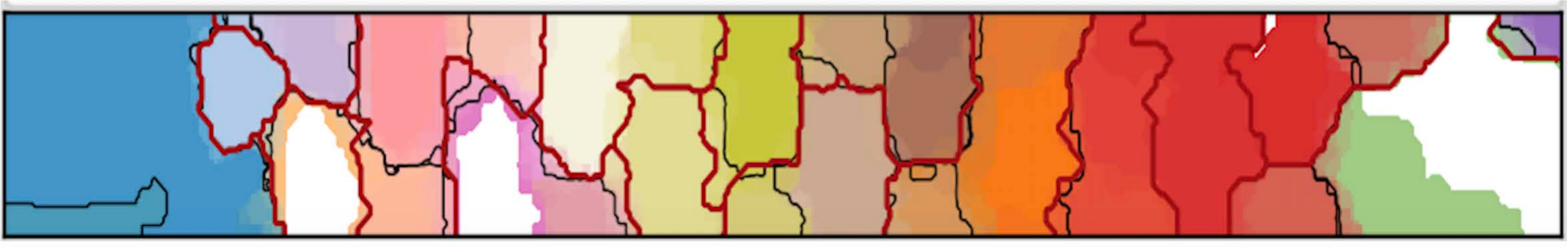}
    \caption{The comparison of the red 2-cell boundaries of Morse complexes that arise from the mean field (red boundaries) with the black expected boundaries, which correspond to points with $\PMap_i = 0.5$ for some cluster $i$.}
    \label{fig:meanVsExpectedFlow}
\end{figure}

Having identified mandatory maximum clusters, we compute the cluster membership probabilities for each position $x \in \Mspace$ and visualize the probabilities through color blending; see Fig.~\ref{fig:approach2Flow}a for the {\PM}. Fig.~\ref{fig:pdfQueryFlow1} visualizes interactive queries of $\PMap$ in the regions with uncertainty at eight query locations to gain a quick insight into uncertainty surrounding those locations. 
The red contours in Fig.~\ref{fig:meanVsExpectedFlow} visualize the spatial inconsistency of 2-cell boundaries of the mean-field Morse complex with the expected boundaries represented by the black contours. 

\begin{figure}[!ht]
  \centering
    \includegraphics[width=0.47\textwidth]{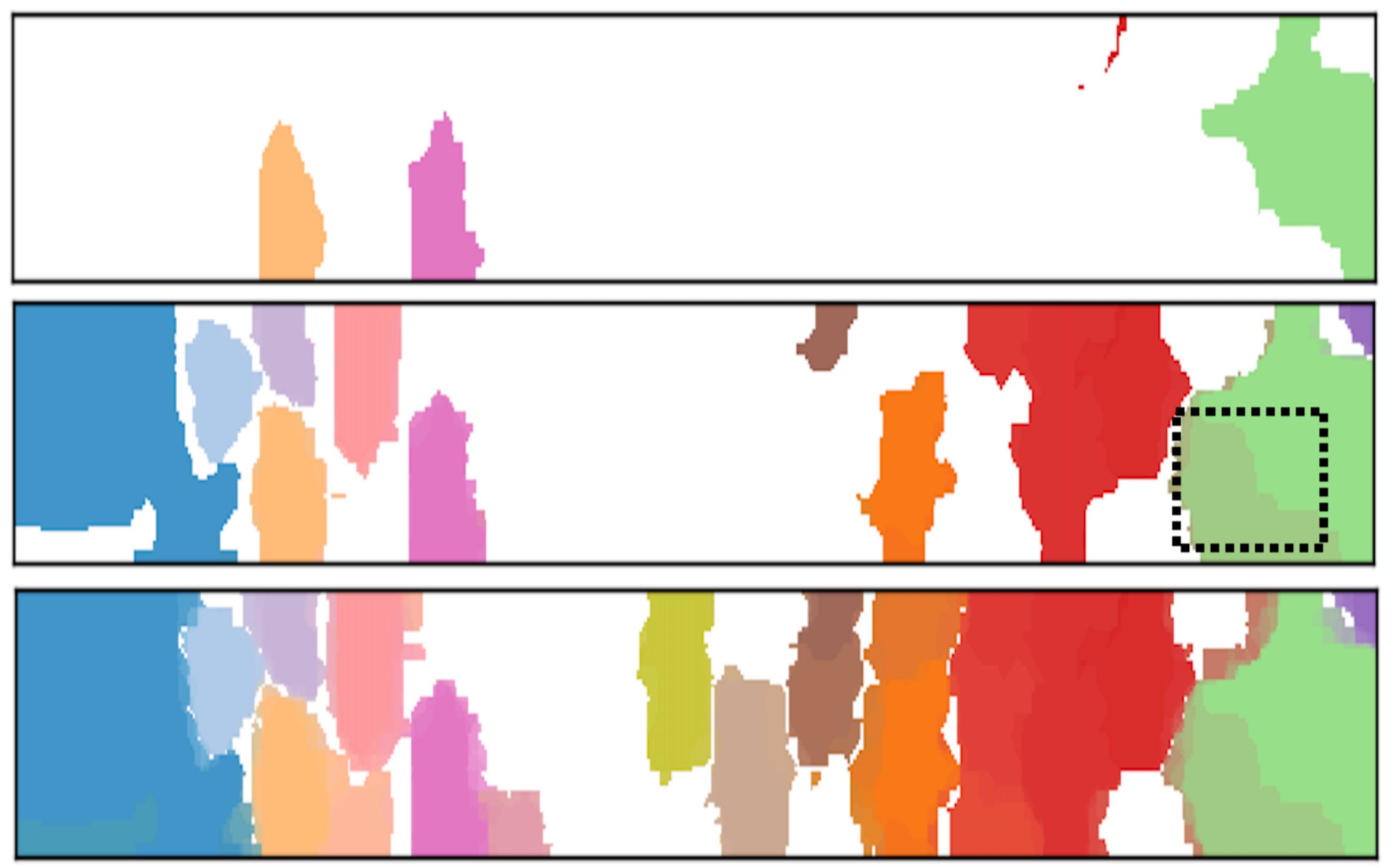}
    \caption{Uncertainty-aware exploration of Morse complex 2-cells for an ensemble of flow simulations across the agreement threshold (from top to bottom) of $95\%$, $80\%$, and $60\%$, respectively.}
    \label{fig:exploreMorseCells}
\end{figure}

For uncertainty-aware exploration of ensemble members, we further study the positional uncertainty among 2-cells by changing the \emph{agreement threshold} $a$.  
As illustrated in Fig.~\ref{fig:exploreMorseCells} top, the 2-cells visualized at threshold $a = 95\%$ represent the points that flow to a single mandatory maximum among at least $95\%$ of ensemble members. 
For instance, the points contained in a magenta 2-cell flow to a magenta cluster (Fig.~\ref{fig:mcpFlow}b) in at least $95\%$ of the ensemble members. 
For $a=80\%$, the region enclosed by the dotted box (Fig.~\ref{fig:exploreMorseCells} middle) highlights the positional uncertainty as a mixture of green and red mandatory maxima. 
The query selection of region $7$ for the same region in Fig.~\ref{fig:pdfQueryFlow1} shows that the gradient flows toward the green and red clusters (Fig.~\ref{fig:mcpFlow}b) in $95\%$ and $5\%$ of ensemble members, respectively.

\begin{figure}[!ht]
  \centering
    \includegraphics[width=0.48\textwidth]{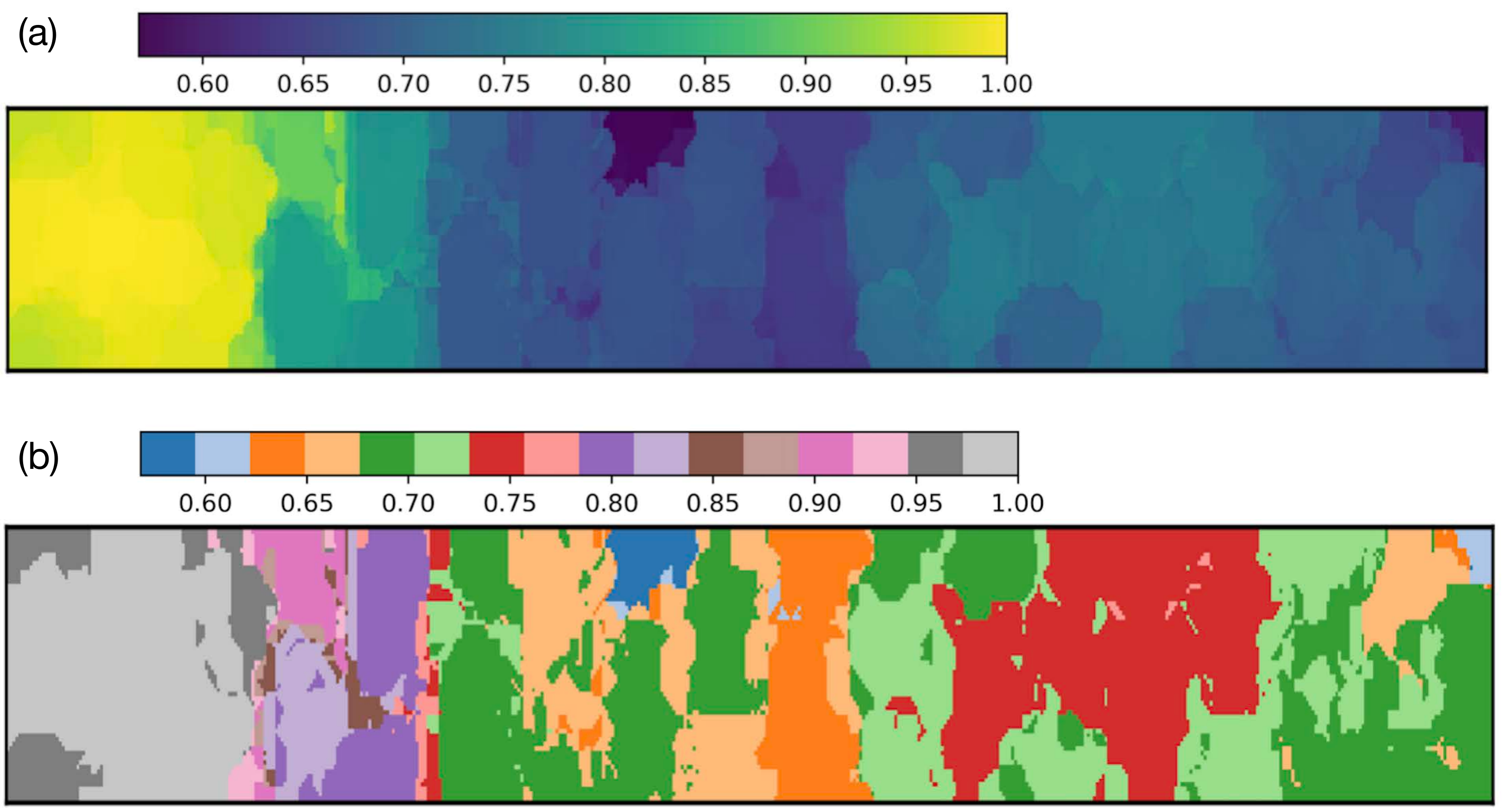}
    \caption{(a) A visualization of the {\SM} for an ensemble of K\'{a}rm\'{a}ns Vortex Street dataset. (b) A quantized visualization of the {\SM}.}
    \label{fig:approach1Flow}
\end{figure}

Finally, we visualize the {\SM} $\SMap$ (Fig.~\ref{fig:approach1Flow}a).
 The 2-cells with different shades of yellow, green, and blue indicate the presence of relatively high, moderate, and low local maxima, respectively, across all ensemble members. 
The quantized visualization in Fig.~\ref{fig:approach1Flow}b offers a further insight: the 2-cells with relatively large color fluctuations indicate the positional uncertainty in their boundaries.

\subsection{Weather Dataset} \label{sec:weather} 

\begin{figure}[!ht]
  \centering
    \includegraphics[width=0.48\textwidth]{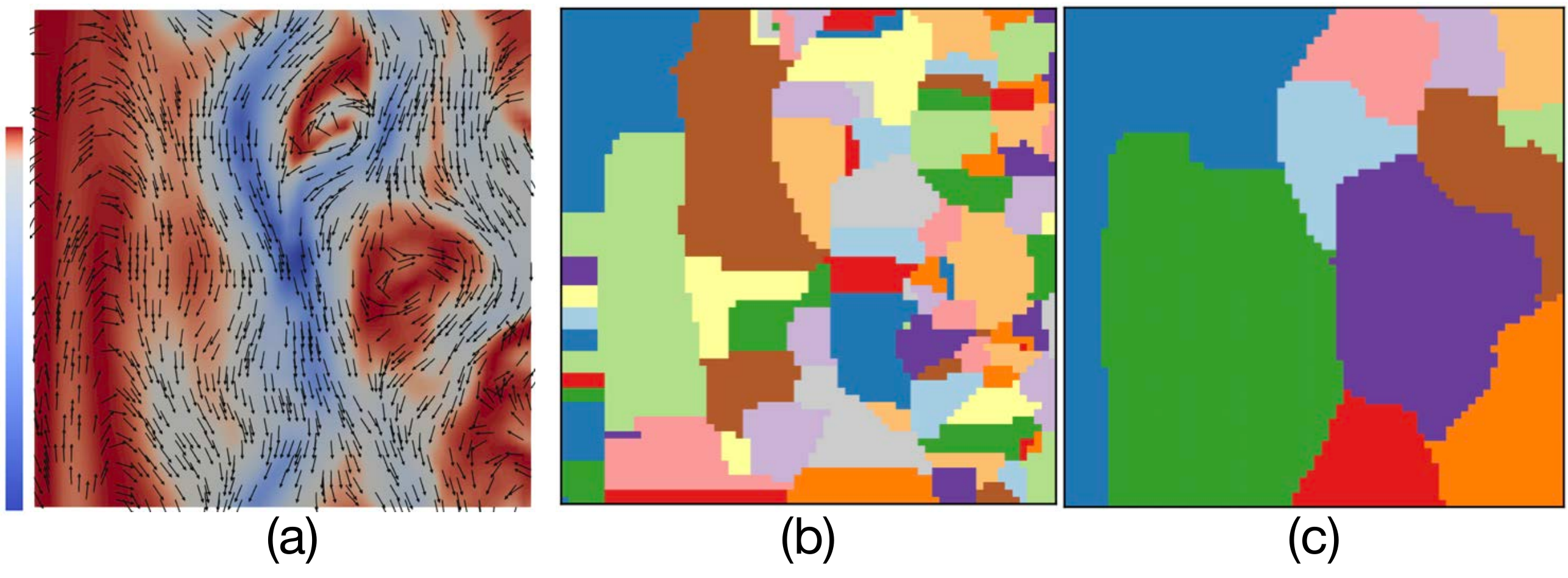}
    \vspace{-4mm}
    \caption{The weather dataset. (a) Mean vector field colored by vector magnitude (the red being high and the blue being low). Morse complexes of the mean field before (b) and after (c) persistence simplification.}
    \label{fig:weather-data}
\end{figure}

\begin{figure}[!ht]
  \centering
    \includegraphics[width=0.48\textwidth]{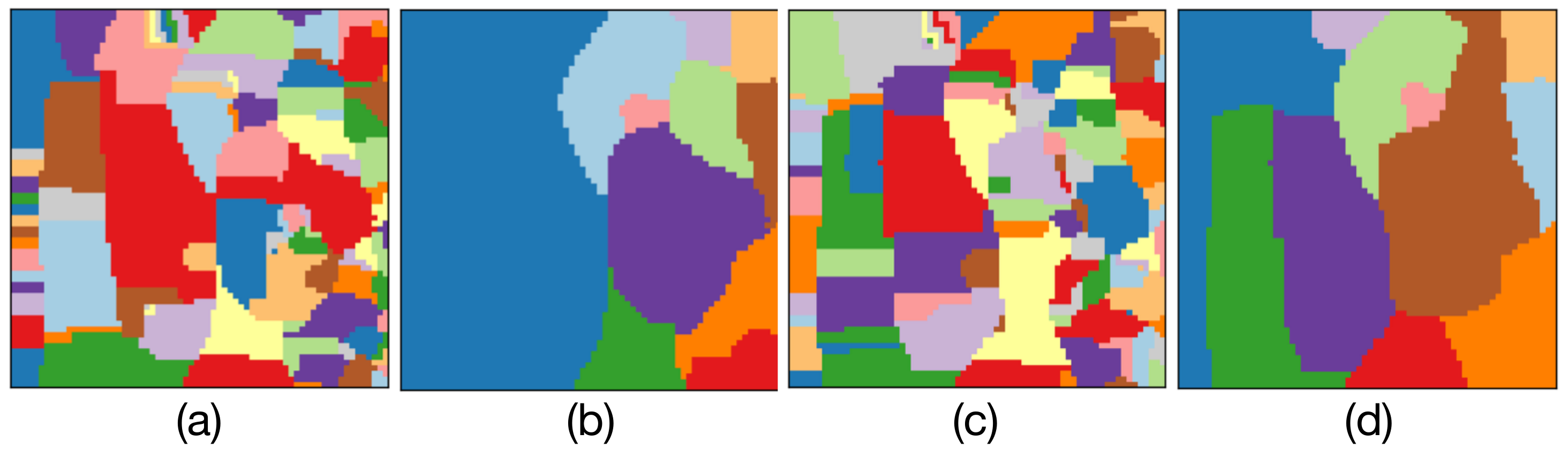}
    \vspace{-4mm}
    \caption{Morse complexes of ensemble members $f_1$ and $f_2$ before (a, c) and after (b, d) persistence simplification.}
    \label{fig:newFlowMorseComplexes}
\end{figure}

In our second real-world example, we analyze an ensemble of vector fields with $15$ members that is part of a climate dataset\footnote{\url{https://iridl.ldeo.columbia.edu/}}. 
Fig.~\ref{fig:weather-data}a shows the mean vector field with color encoding vector magnitudes. Fig.~\ref{fig:weather-data}b and Fig.~\ref{fig:weather-data}c visualize the Morse complex 2-cells before and after persistence simplification of the mean field, respectively. 
Fig.~\ref{fig:newFlowMorseComplexes} shows the Morse complexes for two ensemble members. Their 2-cell boundaries are shown to vary substantially before and after persistence simplification. 

\begin{figure}[!ht]
  \centering
    \includegraphics[width=0.35\textwidth]{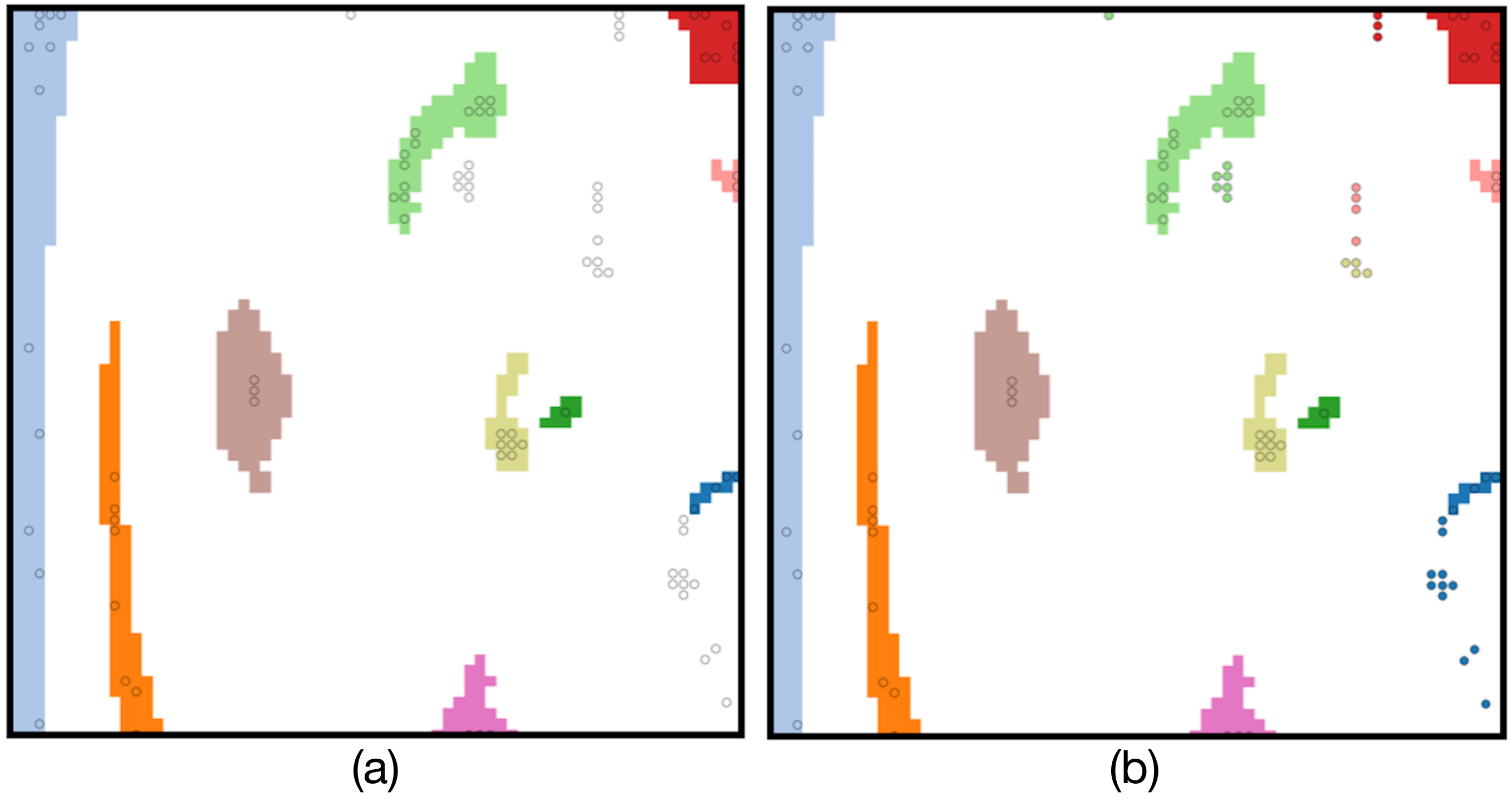}
    \vspace{-4mm}
    \caption{(a) Mandatory local maxima (colored regions) of the weather dataset. Hollow circles represent local maxima of each ensemble member. (b) Each circle is assigned the label of its nearest mandatory local maximum.}
    \label{fig:newFlowMandatoryMaxima}
\end{figure}

\begin{figure}[!ht]
  \centering
    \includegraphics[width=0.49\textwidth]{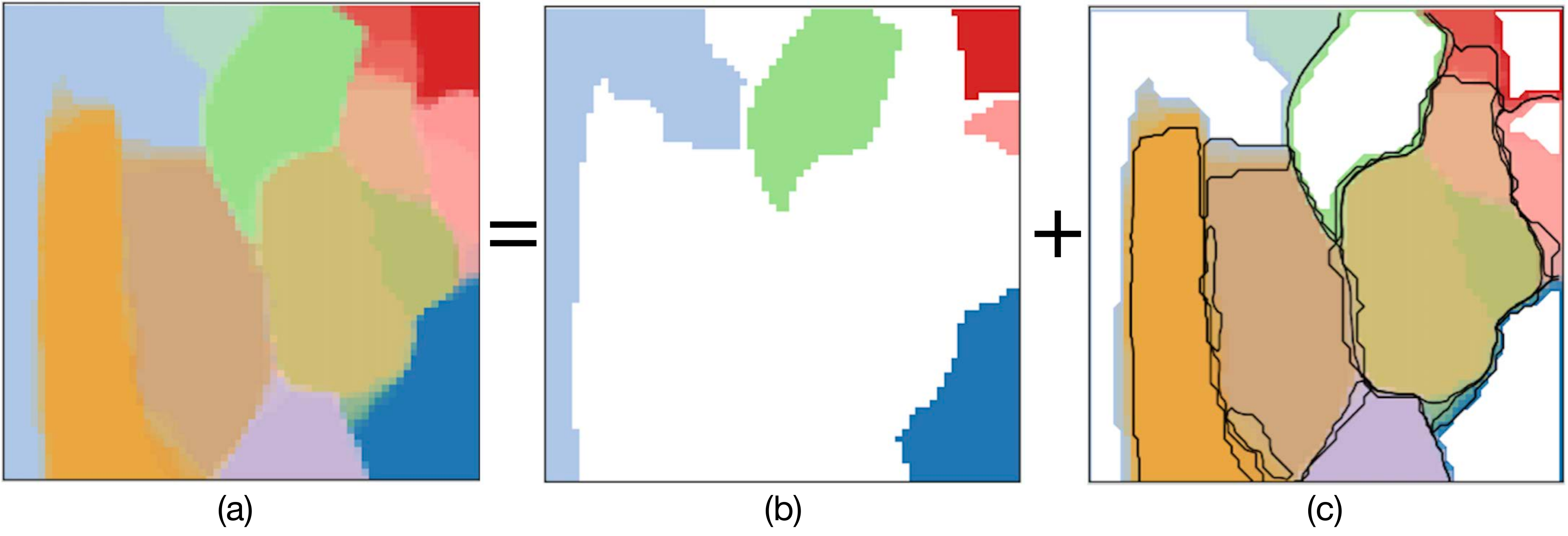}
    \vspace{-4mm}
    \caption{(a) Visualization of the Probabilistic Map for the weather dataset, including (b) points with certainty and (c) points with uncertainty with expected Morse complex boundaries (black contours).}
    \label{fig:probabilisticMapFlow2}
\end{figure}

We first explore the {\PM} $\PMap$ in Fig.~\ref{fig:newFlowMandatoryMaxima}a, which visualizes 10 unique mandatory local maxima of the ensemble. 
Based on our noise model, we simplify the Morse complex for each ensemble member until $10$ local maxima are left. 
For each ensemble member after simplification, we overlay its local maxima (hollow circles) with the mandatory local maxima (colored regions) in Fig. \ref{fig:newFlowMandatoryMaxima}a. We assign labels (colors) to the (hollow) local maxima based on the labeling of their nearest mandatory local maxima in Fig. \ref{fig:newFlowMandatoryMaxima}b. 
We compute the cluster membership probabilities for each domain position and visualize the probabilities through color blending in Fig. \ref{fig:probabilisticMapFlow2}. 

\begin{figure}[!ht]
  \centering
    \includegraphics[width=0.49\textwidth]{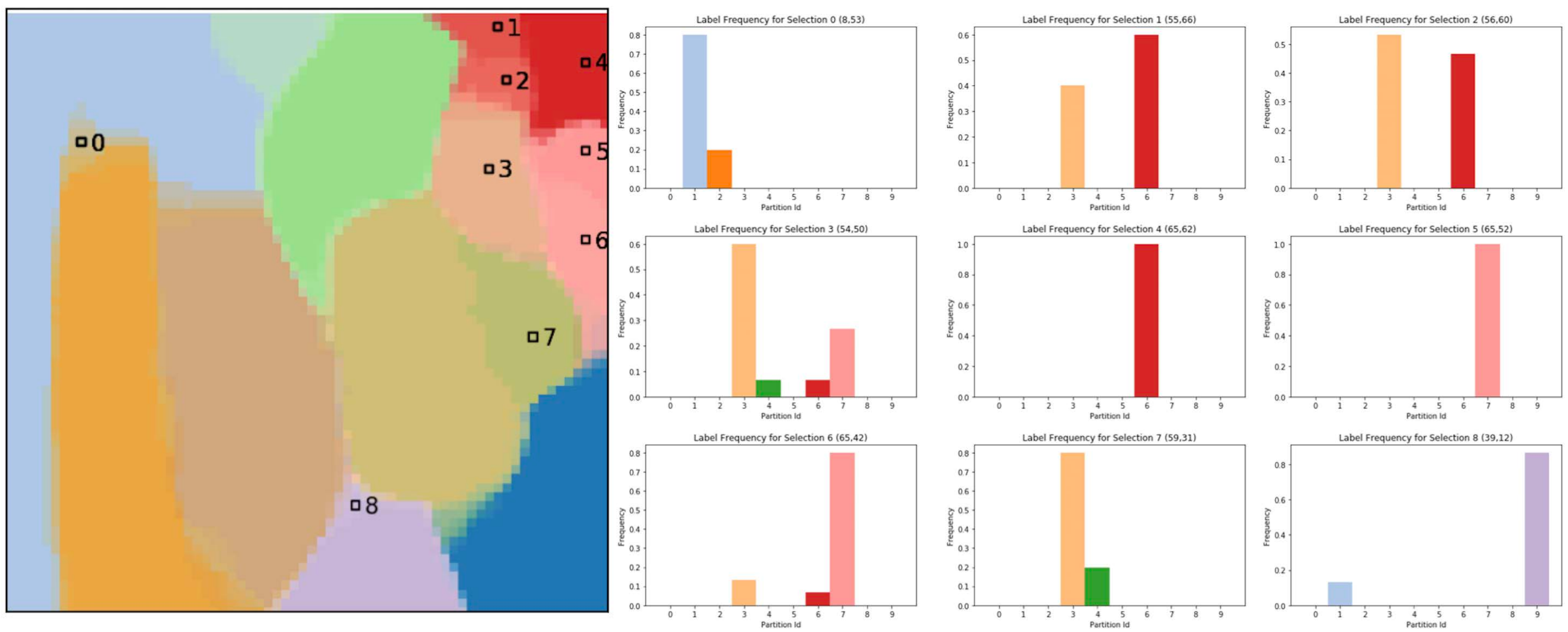}
    \vspace{-4mm}
    \caption{Interactive queries for the uncertain regions of the {\PM} for the weather dataset.}
    \label{fig:pdfQueryFlow2}
\end{figure}

\begin{figure}[!ht]
  \centering
    \includegraphics[width=0.49\textwidth]{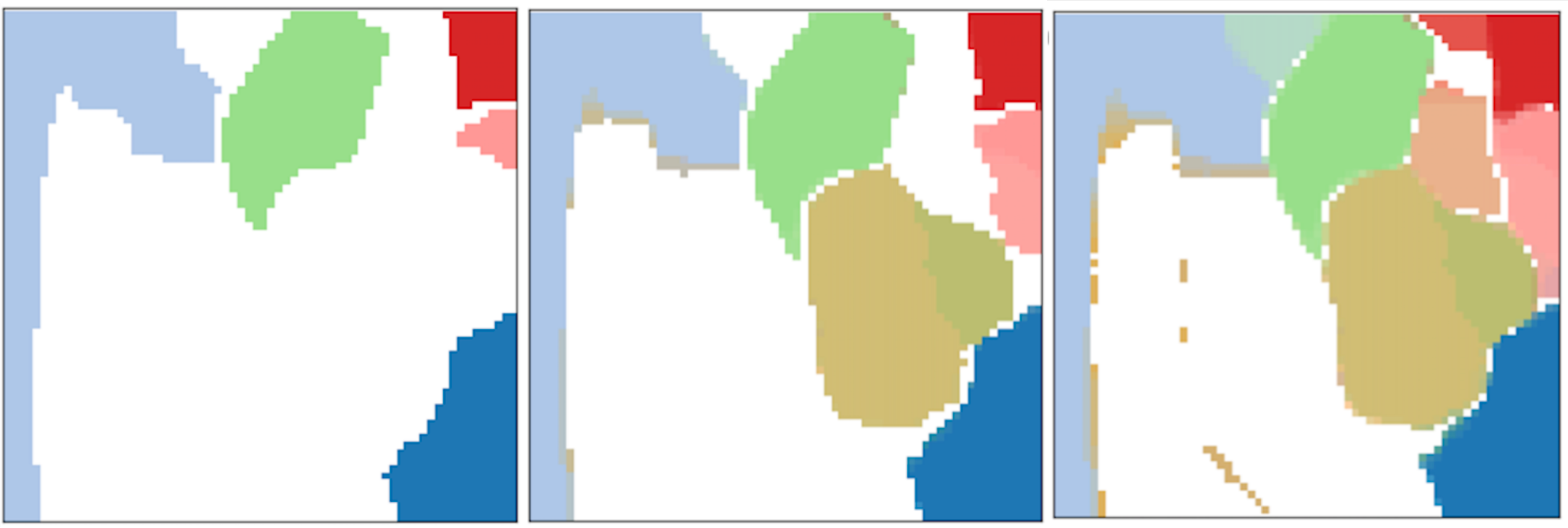}
    \caption{Uncertainty-aware exploration of Morse complex 2-cells for the weather dataset across agreement thresholds (from left to right) of $95\%$, $80\%$ and $60\%$, respectively.}
    \label{fig:exploreVectorFieldMorseCells}    
\end{figure}

Fig.~\ref{fig:pdfQueryFlow2} further visualizes interactive queries of $\PMap$ in the regions with uncertainty. The positional uncertainty among the 2-cells is explored by changing the agreement threshold for ensemble members, as shown in Fig.~\ref{fig:exploreVectorFieldMorseCells}.

\begin{figure}[!h]
  \centering
    \includegraphics[width=0.49\textwidth]{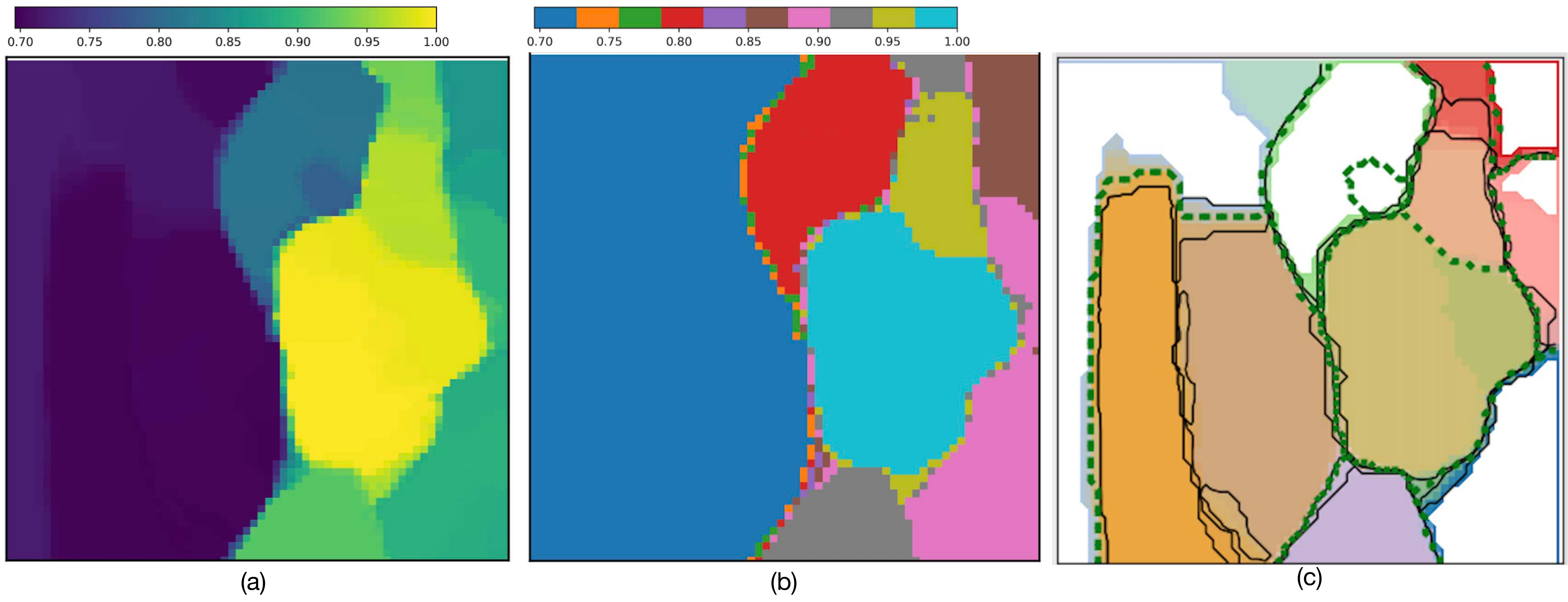}
    \caption{(a)The {\SM} for the weather dataset. (b) Quantized visualization. (c) Mean-field boundaries (dotted green) and expected (black) boundaries for the {\PM} in Fig.~\ref{fig:probabilisticMapFlow2}.}
    \label{fig:survivalMapNewFlow}
\end{figure}

Finally, we compute and visualize the {\SM} $\SMap$ in Fig.~\ref{fig:survivalMapNewFlow}a, which gives insight into the relative heights of local maxima encoded by the survival measure. 
The quantized visualization in Fig.~\ref{fig:survivalMapNewFlow}b gives further insight into the 2-cells of the Morse complex,  similarly to the setting of the K\'{a}rm\'{a}n vortex street dataset.

%% file: sec-methods-noise.tex
\section{Noise Model}
\label{sec:noiseModel}

We justify our noise model with respect to mandatory local maxima~\cite{DavidJosephJulien2014}. 
Specifically, we demonstrate by experiments that mandatory local maxima have a one-to-one correspondence with the local maxima of the ground truth function under our noise model.  

Suppose $n$ ensemble members are given as scalar functions defined on a shared 2D domain, $f_1, \cdots, f_n:  \Mspace \to \Rspace$, where $ \Mspace \subset \Rspace^2$. We study an ensemble of Morse complexes $\MS_1, \cdots, \MS_n$ computed from these functions.  
We assume that each ensemble member $f_i$ is drawn from a  \emph{distribution} that is concentrated around a ground truth function $f$, i.e., $f_i(x) \sim f(x) \pm  \epsilon_i(x)$ for any $x \in \Mspace$. 
Let $p_f$ denote the \emph{persistence} of the smallest topological feature of the ground truth function $f$.
We assume $\epsilon_i(x) < \frac{p_f}{2}$.

For simplicity, for a fixed ensemble member $f_i$, we assume we have uniform noise $\epsilon_i(x) = \epsilon < \frac{1}{2} p_f$, and let $f^+(x) = f(x) + \epsilon$ and $f^-(x) = f(x) - \epsilon$; we have $f^-(x) \leq f_i(x) \leq f^+(x)$ for all $x$. 
Let $D_f$ represents the persistence diagram of the sublevel set filtration of $f$. 
Based on the stability of the persistence diagrams~\cite{CohEdeHar2007}, 
$d_B(D_f, D_{f_i}) \leq ||f - f_i||_\infty \leq \epsilon$. 
However, the stability of critical values is not the same as the stability of critical points; therefore, we need additional machinery. 
Using the stability of critical points with interval persistence~\cite{DeyWenger2007}, we can show that under our noise model, there is no pairing switches among the critical points;  therefore, there is a one-to-one correspondence between the mandatory local maxima and the local maxima of the ground truth function.

\begin{figure}[!ht]
  \centering
    \includegraphics[width=0.49\textwidth]{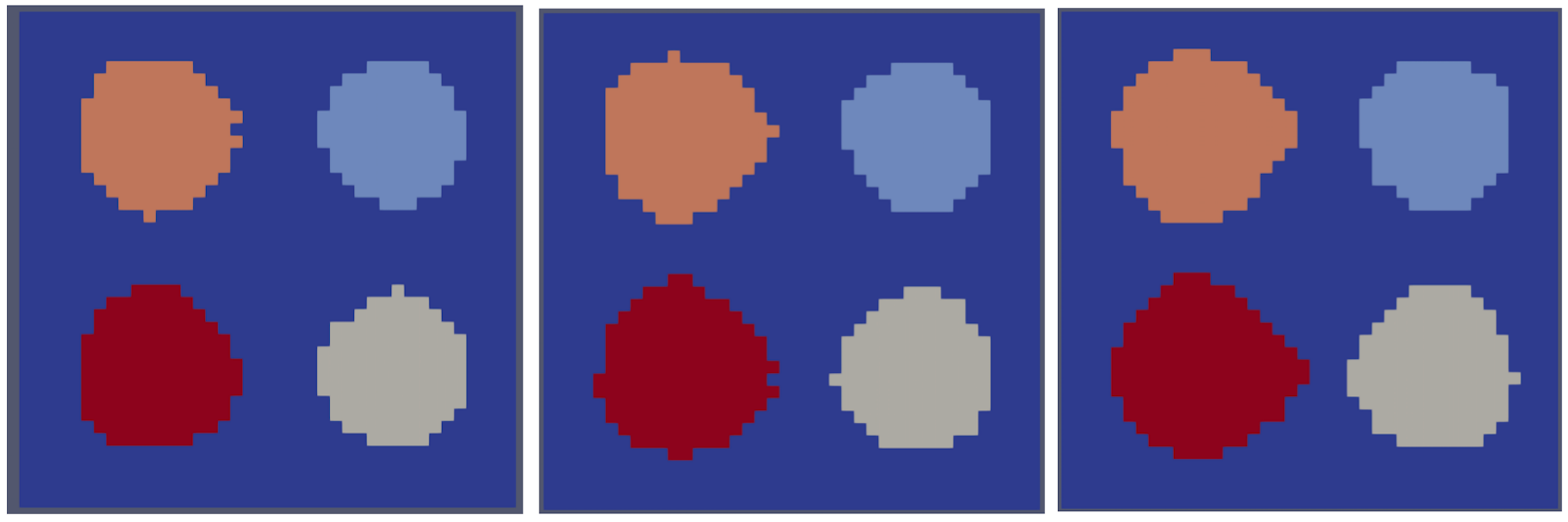}
    \caption{Visualizing mandatory local maxima for three ensembles that arise from a mixture of four Gaussians corrupted by noise. The mandatory local maxima are visualized for three noise levels: $\epsilon = 0.95 \times \frac{p_f}{2}$, $\frac{p_f}{2}$, and $1.05 \times \frac{p_f}{2}$. The number of mandatory maxima is equal to the number of local maxima in the ground truth when $\epsilon < p_f/2$. With $\epsilon \ge p_f/2$, the mandatory local maximum clusters start expanding and eventually merge with one another.}
    \label{fig:mcp-relation}
\end{figure}

We demonstrate the above one-to-one correspondence via a synthetic dataset that arises from a mixture of four Gaussians, $f = \sum_{i=1}^{4} \mathcal{N}(\mu_i,\,\sigma_i^{2})$. 
We compute $p_f$ for the ground truth function $f$. 
Three synthetic ensemble datasets are then generated with three noise levels,
$\epsilon = 0.95 \times \frac{p_f}{2}$, $\frac{p_f}{2}$, and $1.05 \times \frac{p_f}{2}$.  
The mandatory maxima are computed for each ensemble. 
In all cases with $\epsilon < \frac{p_f}{2}$, we obtain the same number (four) of mandatory local maxima as the number of local maxima in the ground truth. Fig. \ref{fig:mcp-relation} illustrates the result for one such experiment; the visualization is generated using the topology toolkit \cite{TA:2018:TFL}. For $\epsilon < p_f/2$, we also get clear separation among mandatory local maxima. With $\epsilon \ge p_f/2$, the mandatory maxima start expanding and eventually merge with one another.

\begin{figure}[!ht]
  \centering
    \includegraphics[width=0.4\textwidth]{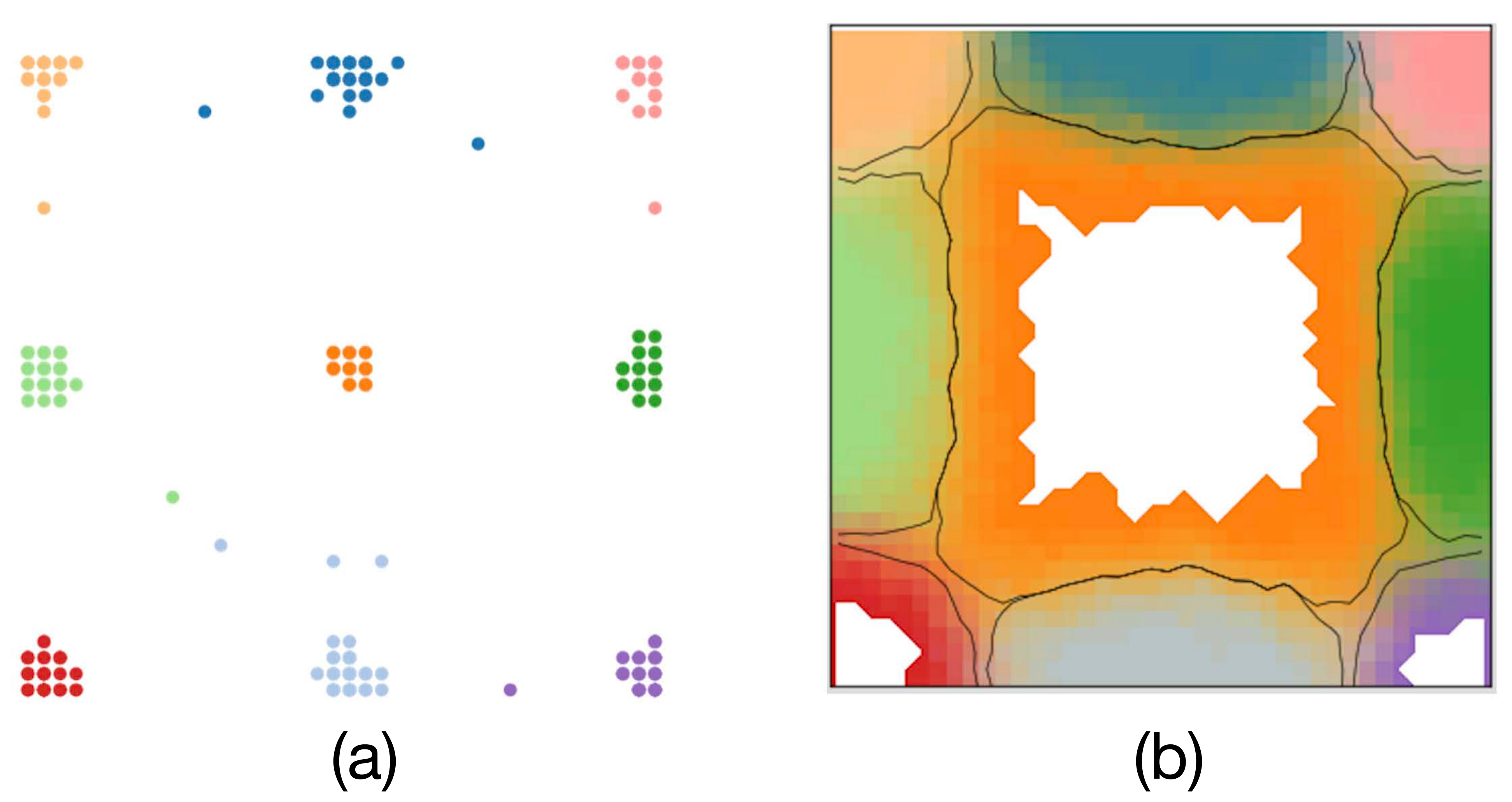}
    \vspace{-4mm}
    \caption{{\PM} in (b) derived using the local maxima clustering visualized in (a) for $\epsilon = p_f$.}
    \label{fig:brokenNoiseAssumption}
\end{figure}

\para{Violation of our noise model.} 
One-to-one correspondence (between mandatory critical points and critical points of all ensemble members) may not be preserved for certain ensemble members that do not conform to our noise model (i.e., $\epsilon > p_f/2$), we would need to employ heuristics to deal with such cases; this include computing the nearest mandatory maxima and clustering of local maxima. 
For the real-world datasets (Sec.~\ref{sec:karmanVortexStreet} and Sec.~\ref{sec:weather}), we implement the nearest mandatory maximum heuristics (Fig.~\ref{fig:mcpFlow} and Fig.~\ref{fig:newFlowMandatoryMaxima}) to derive the \PM.
The clustering of local maxima is illustrated in Fig.~\ref{fig:brokenNoiseAssumption}a for deriving the {\PM} (Fig.~\ref{fig:brokenNoiseAssumption}b) under a noise level $\epsilon = p_f$. 

\begin{figure}[!h]
\centering 
\includegraphics[width=0.80\columnwidth]{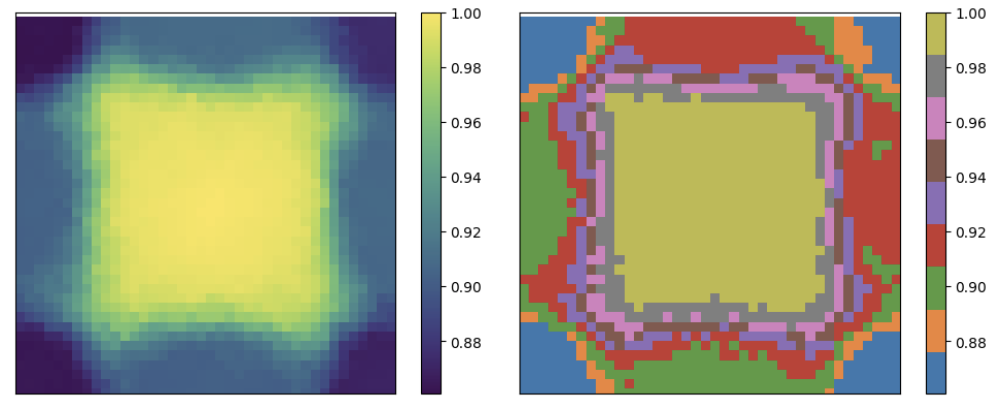}
\vspace{-2mm}
\caption{The {\SM} $\SMap$ (left) and quantized visualizations (right) of the Ackley dataset for noise level $\epsilon = 1.5 \times \frac{p_f}{2}$.}
\label{fig:higherNoiseSurvivalMapAckley}
\end{figure}

Furthermore, for quantized visualization, the color fluctuations spatially grow in size with increase in the noise level (that violates our noise model), as illustrated in Fig.~\ref{fig:higherNoiseSurvivalMapAckley}}.

%% file: sec-conclusion.tex
\section{Conclusion and Future Work}
\label{sec:conclusion}

We study Morse complexes for ensembles representing uncertain 2D scalar data. 
We propose statistical summary maps as new abstractions for quantifying structural variations among ensembles of Morse complexes. 
We introduce two types of statistical summary maps, the {\PM} $\PMap$  and the {\SM} $\SMap$. 
The {\PM} takes advantage of mandatory maxima~\cite{DavidJosephJulien2014} whereas the {\SM} leverages local gradient flows based on persistence simplification~\cite{EdelsbrunnerLetscherZomorodian2002}.
We employ uncertainty visualization methods such as color mapping, interactive distribution queries, and uncertainty-aware exploration to understand the structural variability captured by our statistical summary maps.

For future work, we plan to generalize our noise model. 
We also would like to extend our work for Morse complexes beyond 2D. 
While Morse complexes may be approximated in higher dimensions, visualizing positional uncertainties in higher dimensions will require new visual mappings.